\documentclass[%
 reprint,
superscriptaddress,
 amsmath,amssymb,
 aps,longbibliographystyle
float,
floatfix,
]{revtex4-2}

\usepackage{graphicx}
\graphicspath{{images/}}
\usepackage{dcolumn}
\usepackage{bm}
\usepackage{physics}

\usepackage{array,multirow}

\usepackage[utf8]{inputenc}
\usepackage{upgreek}

\usepackage{mathrsfs}
\usepackage{amsmath}
\usepackage{amsfonts}
\usepackage{bbm}

\usepackage{tikz}
\usetikzlibrary{positioning,shapes}

\usepackage{hyperref}
\hypersetup{colorlinks=true,
citecolor=blue,
linkcolor=brown,
urlcolor=blue,
}

\definecolor{johnsgreen}{rgb}{0.09, 0.65, 0.27}

\newcommand{\QSERG}{Quantum Systems Engineering Research Group, Department 
of Physics, Loughborough University, Leicestershire LE11 3TU, United Kingdom}

\begin{document}


\title{Generalized Phase-Space Techniques to Explore Quantum Phase Transitions 
in Critical Quantum Spin Systems}

\author{N.~M.~Millen}
\affiliation{\QSERG}

\author{R.~P.~Rundle}
\affiliation{School of Mathematics, Fry Building, University of Bristol, UK}
\affiliation{\QSERG}
\affiliation{The Wolfson School, Loughborough University, Loughborough, Leicestershire LE11 3TU, United Kingdom}

\author{J.~H.~Samson}
\affiliation{\QSERG}

\author{Todd Tilma}
\affiliation{Department of Physics, College of Science, Tokyo Institute 
of Technology, H-63, 2-12-1 \=Ookayama, Meguro-ku, Tokyo 152-8550, Japan}
\affiliation{Quantum Computing Unit, Institute of Innovative Research, 
Tokyo Institute of Technology, S1-16, 4259 Nagatsuta-cho, Midori-ku, 
Yokohama 226-8503, Japan}
\affiliation{\QSERG}

\author{R.~F.~Bishop}
\email{raymond.bishop@manchester.ac.uk}
\affiliation{Department of Physics and Astronomy, Schuster Building, 
The University of Manchester, Manchester, M13 9PL, United Kingdom}
\affiliation{\QSERG}

\author{M.~J.~Everitt}
\email{m.j.everitt@physics.org}
\affiliation{\QSERG}

\date{\today}

\begin{abstract}
We apply the generalized Wigner function formalism to detect and 
characterize a range of quantum phase transitions in several cyclic, finite-length, 
spin-$\frac{1}{2}$ one-dimensional spin-chain models, viz., the Ising and 
anisotropic $XY$ models in a transverse field, and the $XXZ$ 
anisotropic Heisenberg model.
We make use of the finite system size to provide an 
exhaustive exploration of each system's single-site, bipartite 
and multi-partite correlation functions. In turn, we are 
able to demonstrate the utility of phase-space techniques in 
witnessing and characterizing first-, second- and 
infinite-order quantum phase transitions, while also enabling 
an in-depth analysis of the correlations present within critical systems. 
We also highlight the method's ability to capture other features of 
spin systems such as ground-state factorization and critical system scaling.
Finally, we demonstrate the generalized Wigner function's utility for state
verification by determining the state of each system and their constituent
sub-systems at points of interest across the quantum phase transitions, 
enabling interesting features of critical systems to be intuitively 
analyzed. 
\end{abstract}

\maketitle

\section{\label{Intro}Introduction}

Quantum many-body systems often present interesting and unexpected 
properties that have no classical counterparts, and which 
are ultimately due to the strongly correlated and
highly entangled nature of the underlying many-body states. 
Quantum spin-lattice systems in particular are able to elucidate 
some of the underlying properties and characteristics of 
more general many-body systems by providing a simple yet 
rich and valuable framework that constitutes the
basic models of many physical systems such as magnetic insulators 
\cite{QuantumMagnetism-book,SpinChain1B,SpinInfiniteLattice} 
and coupled qubits \cite{QI_2}. The study of spin-lattice models
of quantum magnetism has thus become an area of enormous theoretical
study in recent years.  Additional impetus has come from the 
growing realizations that quantum many-body systems themselves 
provide valuable resources for quantum computing (see, e.g., 
Refs.~\cite{QC_1,QC_2,QI_2}), and that quantum information 
theory concepts can themselves provide valuable insights into 
the properties and behaviors of quantum many-body systems, 
typically via a detailed study of the intimate entanglement 
structure of many-body wave functions~\cite{QI_1}.

From a modern perspective quantum entanglement \cite{EntanglementMeasuresRev} 
has become a key element of various quantum technologies, due to
the significance of entangled states as a resource in applications such
as the development of protocols for teleportation, secure quantum 
key distribution schemes, etc.  Means to produce various entangled
states both systematically and efficiently are hence of growing
importance.  For example, Einstein-Podolsky-Rosen (EPR) pairs, which
may be considered as qubit pairs in the form of a Bell state,
furnish one of the simplest illustrations of maximum
entanglement at the bipartite level.  The demonstration that such 
EPR pairs can be rapidly and robustly generated with spin chains \cite{QI1}
has also provided additional impetus for their study.

One especially interesting feature of spin system models is their 
ability to exhibit and transition between different quantum phases
\cite{QPTMethod4,QPTMethod1}. Such phase transitions occur in
interacting many-body ensembles in the thermodynamic limit.
The more well-known thermal phase transitions that occur at specific 
critical temperatures $T_c$ are due to thermal fluctuations that 
represent the competition between the entropy
and the internal energy of the system, which themselves depend on 
such parameters as the pressure ($P$), volume ($V$) and temperature
($T$).  Such thermal phase transitions (in one-component systems)
are typically studied by thermodynamic free energy state functions 
such as the Gibbs free energy, $G(P,T)$ or Helmholtz free energy,
$F(V,T)$, which become doubly degenerate at the corresponding phase 
boundary between two thermal phases.

By contrast, the quantum phase transitions (QPTs) 
\cite{QuantPhaseTrans,QuantPhaseTrans2,InfQPT1} of interest here
occur at zero temperature ($T=0$) between differing ground states 
of the system, and are due to quantum fluctuations that represent 
the competition between at least two non-commuting terms in the system
Hamiltonian, each of which by itself promotes a different form
of ground state. When acting together they therefore act to
{\it frustrate\,} one another.  A prototypical frustrated spin-lattice
system is the spin-$\frac{1}{2}$ $J_1$--$J_2$ model on a one-dimensional (1D)
chain \cite{J1J2QPT1a,J1J2QPT1b} or a two-dimensional square lattice 
\cite{J1J2QPT2a,J1J2QPT2b,J1J2QPT2c}, in which the two competing
terms in the Hamiltonian are isotropic Heisenberg spin-spin interactions
between nearest-neighbor and next-nearest-neighbor pairs, with strength
parameters $J_1 \equiv \lambda_1$ and $J_2 \equiv \lambda_2$,
respectively. 

If the set of variables 
$\{\lambda_1, \lambda_2, \cdots, \lambda_n\}$
represents the strength parameters of the various non-commuting parts in 
the model Hamiltonian, then the ground-state energy of the system,
$E_0(\lambda_1, \lambda_2, \cdots, \lambda_n)$, correspondingly
becomes doubly degenerate at the boundary 
$(\lambda_1^c, \lambda_2^c, \cdots, \lambda_n^c)$
between two quantum phases. Such QPTs are said to be a first-order QPT (1QPT)
or a second-order QPT (2QPT, also known as a continuous QPT) depending, 
respectively, on whether the first-order partial derivatives 
$\{\partial E_0/\partial \lambda_i, i=1,2,\dots,n\}$ are discontinuous
or continuous at the transition boundary.  Whereas a 2QPT is typically characterized 
by the existence of an infinite correlation length in the system and a 
corresponding power-law decay in its correlations, 
which is often expressed via a divergence (or finite discontinuity) in one or more 
second-order derivatives of the ground-state energy, there also exist
infinite-order QPTs ($\infty$QPTs) at which {\it all\,} finite-order
derivatives of the ground-state energy remain continuous. In the 
present work we will consider spin-lattice systems that display all 
three types of QPTs (i.e., 1QPT, 2QPT, $\infty$QPT).

The study of QPTs has itself acquired additional impetus from 
the increasingly widespread use of ultra-cold atoms trapped in optical 
lattices formed by a periodic potential, which itself has been created 
by standing waves formed from a suitable array of lasers, 
in order to simulate a wide variety of of systems of interest 
in condensed matter and many-body physics 
\cite{Duan-Demler-Lukin_2003,ColdAtoms1,ColdAtoms2,QuantCorrelations2}, 
especially spin-lattice systems. In such simulations it is
then often experimentally possible to vary the strength parameters
$\{\lambda_1, \lambda_2, \cdots, \lambda_n\}$ discussed above, and 
hence to map out a QPT and to find its boundary 
$(\lambda_1^c, \lambda_2^c, \cdots, \lambda_n^c)$
in this parameter space.  Examples include that from a superfluid to a Mott
insulator \cite{MOTTQPT1,MOTTQPT2}, as well as  many others in quantum 
magnetism, both for one-dimensional spin chains 
(see, e.g., Ref.~\cite{HTSuper}) and for periodic lattices in two or
more dimensions 
(see, e.g., Refs.~\cite{Duan-Demler-Lukin_2003,Tao-et-al_2014,Dey-Sansarma_2016}).
For example, a two-dimensional regular honeycomb lattice may be formed
by interfering three coplanar laser beams propagating at relative angles of 
$\pm 120^{\circ}$, and Duan {\it et al.} have shown how to engineer 
specific spin Hamiltonians on such a lattice accordingly ~\cite{Duan-Demler-Lukin_2003}.
Similarly, concrete proposals to form optical lattices 
representing honeycomb-lattice
bilayers in both $AA$ stacking~\cite{Tao-et-al_2014} and $AB$ (or Bernal) 
stacking~\cite{Dey-Sansarma_2016}, both of which use five lasers,
have also been discussed.

Exact solutions for the ground-state energy of quantum many-body
systems in general \cite{Condensed}, and spin-lattice systems in 
particular \cite{SpinChain1B}, are known only in a few special cases. 
Otherwise, one has to resort to various tools of quantum many-body
theory to find approximate solutions.
Furthermore, witnessing critical behavior in these systems, even 
when some exact results are known, can be quite 
difficult and cumbersome (see, e.g., Refs.~\cite{QPTMethod1, QPTMethod2}). 
However, the crucial role that quantum correlations play in critical many-body systems 
has highlighted the use of entanglement and correlation measures 
as possible methods for witnessing and characterizing 
QPTs \cite{QCaQI,XXZ_QPT_4}. The subsequent application of quantum 
information tools to critical quantum systems has by now firmly demonstrated 
their unparalleled ability to measure and characterize quantum critical 
behavior in a wide range of many-body systems 
\cite{XY_QPT_4,XY_QPT_6,QuantCorrelations6,QPTEntanglement7,QPTEntanglement5,QPTEntanglement3,QPTSpin1,QPTEntanglement4,QuantCorrelations4,QuantCorrelations5,QuantCorrelations3}.
    
Within the wider context of the phase-space formulations of
quantum mechanics and the associated quantum distribution functions
(see, e.g., Ref.~\cite{DISTRIBUTIONFUNCTIONSINPHYSICS}),
the Wigner function has found a large number of applications to
a wide variety of problems in physics.  Examples include the calculation
of quantum mechanical observables in quantum ballistic 
transport studies~\cite{Iafrate-Grubin-Ferry_1982} and many
other areas such as quantum optics, quantum electronics, quantum
chemistry, signal processing, and quantum information theory.
Reference~\cite{Wigner2} presents a nice review of some of 
these latter applications.  

More recently, it has been realized that
the Wigner function in particular can also
provide a useful measure of quantum correlations 
\cite{WigEnt_1,WigEnt_2,WignerEntanglement1,WignerEntanglement2}. 
Consequently, it is natural to explore the application of this 
formalism to the study of QPTs and quantum critical behavior.
Extensions of the original Wigner function formalism for continuous 
systems in the usual position-momentum phase space~\cite{Wigner1,OConnell-Wigner_1981,Iafrate-Grubin-Ferry_1982} to discrete, finite-dimensional systems, such as ensembles of spins, has been a challenging endeavor. 

There have been many attempts to create a finite analog to the Wigner 
function~\cite{Wigner3, Wigner4, Wigner5, Wigner6}, each with their own formulation.
Here we follow the formulation of a generalized Wigner function (GWF) introduced in Ref.~\cite{GWF1},
that is able accurately to construct a complete, continuous Wigner function 
for {\it any\,} arbitrary quantum system, as well as its corresponding Weyl function \cite{QuantumAsStatTheory}.
The latter work in particular has completed the phase-space formulation of 
quantum mechanics by extending the original work of 
Wigner \cite{Wigner1}, Weyl \cite{Weyl1927}, Moyal \cite{Moyal1949}, Groenewold \cite{Groenewold1946}, 
and others to {\it any\,} quantum system.

Research into the application of Wigner function formalisms to the study of 
critical quantum spin systems was recently carried out by Mzaouali {\it et al.} \cite{Discrete}. 
They successfully demonstrated the utility of phase-space techniques for 
witnessing, characterizing and distinguishing first-, second- and infinite-order QPTs. 
In addition to this, they demonstrated 
the ability of the method to detect more nuanced features of these systems such as 
ground-state factorization.  In this work we aim both to verify these 
results and to build upon them by applying the GWF formalism to several
finite-sized ($N=6$) spin-$\frac{1}{2}$ one-dimensional (1D) models, viz., 
the Ising  ferromagnetic chain in a transverse magnetic field,
the anisotropic $XY$ chain in a transverse magnetic field, and the $XXZ$ chain. 
We will explore their individual properties and critical behavior by 
providing an exhaustive exploration of the correlations present within 
these systems.  
We will further demonstrate the ability of the GWF fomalism to 
witness and characterize first-, second- and infinite-order QPTs, as well 
as ground-state factorization, through single, bipartite and multi-partite 
correlation functions. In addition, we demonstrate the formalism's 
ability to capture scaling of finite spin chain systems. We also employ 
the GWF's ability for intuitive state analysis \cite{Wigner7} by visualizing
the state of the system and its constituent 
sub-systems at points of interest across QPTs through spin Wigner function plot 
visualizations. In turn, we demonstrate the utility of this visualization technique 
specifically to witness features of infinite-order QPTs and ground-state factorization.

The remainder of this paper is organized as follows. In Sec.~\ref{Wigner} we 
outline the GWF formalism and discuss its 
application to spin systems. We then apply this formalism in 
Sec.~\ref{SpinChains} to the three spin-$\frac{1}{2}$ chain models
of interest here that we have cited above.  For each model we  produce a phase line 
plot to witness critical behavior in the system and then determine 
the state of the system at points of interest across the phase line 
through the use of spin Wigner function visualizations and 
reference state plots. Finally, in Sec.~\ref{conclusion} we conclude 
and summarize our findings.

\section{\label{Wigner}The Wigner Function}

We make use of a continuous, informationally complete, spin Wigner function~\cite{GWF1, GWF2, Wigner2} with similar properties to the one originally formulated by Wigner in 
1932~\cite{Wigner1}. For a two-level quantum system it takes the form of a quasi-probability distribution defined over the same Euler angles as the Bloch sphere. 
The spin-Wigner function can be expressed in terms of the expectation value of an analog to the usual displaced parity-operator, given by
\begin{equation}
    \hat{\Delta}(\theta,\phi)\equiv\hat{R}(\theta, \phi, \Phi)\hat{\Pi}\hat{R}^{\dagger}(\theta, \phi, \Phi)
    \,,
    \label{Kernel1}
\end{equation}
where the Euler angles $\theta \in [0, \pi]$ 
and $\phi \in [0, 2\pi)$ 
are set by the rotation operator $\hat{R}(\theta, \phi,\Phi)\equiv\mathrm{e}^{-i\hat{\sigma}^{z}\phi/2}\mathrm{e}^{-i\hat{\sigma}^{y}\theta/2}
\mathrm{e}^{-i\hat{\sigma}^{z}\Phi/2}$ and parametrize 
the phase space, and $\hat\Pi \equiv \frac12\left(\hat{\mathbbm 1} + \sqrt 3 \hat \sigma^z\right)$ 
\footnote{Note that $\Phi$ does not contribute to the operator $\hat{\Delta}(\theta,\phi)$ since $\hat{\sigma}_z$ and $\hat{\Pi}$ commute}. 
For a composite 
system of qubits the corresponding displaced parity operator is simply the tensor product, 
\begin{equation}
    \hat{\Delta}(\boldsymbol{\theta},\boldsymbol{\phi})=\bigotimes_{i=1}^N
    \hat{\Delta}(\theta_i,\phi_i) \,,
    \label{Kernel2}
\end{equation}
where $N$ is the number of spins.
The Wigner function,
\begin{equation}\label{theWignerFunction}
    W(\boldsymbol{\theta},\boldsymbol{\phi}) = \Tr\left[\hat\rho \hat{\Delta}(\boldsymbol{\theta},\boldsymbol{\phi})\right]\,,
\end{equation}
where $\hat{\rho}$ is the total density matrix of the spin system,
is a function on the product of $N$ spheres. For the purposes of visualization 
(see, for example, Fig.~\ref{TISpinPlots}) it often suffices to restrict to the \emph{equal-angle} slice;
\begin{widetext}
\begin{equation}
    W^{\mathrm{EA}}(\theta,\phi)=\Tr \left[\hat\rho
    \hat{\Delta}(\theta_1=\theta_2=\ldots=\theta,\phi_1=\phi_2=\ldots=\phi)
    \right] \,.
    \label{WEA}
\end{equation}
In this representation, high-order spherical harmonics are then a witness 
to many-spin entanglement (see, for example, Fig.~7 of Ref.~\cite{Wigner7}).
Following Ref.~\cite{Discrete} we will also use, as a witness of phase transitions, just one specific point 
of the Wigner function equal-angle slice, $W^{\mathrm{EA}}(0,0)$ (see, for example, Fig.~\ref{TIPhaseLine}).

The GWF is our tool to visualize the density matrix $\hat \rho$ (or $\hat \rho_{\mathrm{tot}}$) of the ground state of the full system. We will also visualize correlations through the corresponding reduced density matrices $\hat{\rho}_I$, where $I\subseteq \mathrm{tot}$ is a subset of the spin labels $\mathrm{tot} = \{i:1\le i \le N\}$. 
The reduced density matrix is the partial trace of the density matrix over the remaining qubits $\mathrm{tot}\setminus I$. The corresponding reduced Wigner function is the marginal obtained by integrating the full Wigner function over the angular variables of the remaining qubits. The correspondence follows from repeated use of
\begin{equation}
\frac{1}{2\pi}\int_0^{\pi}\sin\theta_i \; \mathrm{d} \theta_i 
    \int_0^{2\pi} \mathrm{d}\phi_i\hat{\Delta}(\boldsymbol{\theta},\boldsymbol{\phi})=\left(\bigotimes_{j=1}^{i-1}
    \hat{\Delta}(\theta_j,\phi_j)\right) \otimes \hat{\mathbbm 1} \otimes \left(\bigotimes_{k=i+1}^N
    \hat{\Delta}(\theta_k,\phi_k)\right). 
    \label{reduced}
\end{equation}
\end{widetext}

Our choices of reduced Wigner functions for the 6-spin systems studied hare 
are shown in Fig.~\ref{ChainCorrelationKey}, where we label 
$\rho_{\mathrm{tot}} \equiv W^{\mathrm{EA}}(\theta,\phi)$.
We first calculate a reduced Wigner function $W_I(\boldsymbol{\theta},\boldsymbol{\phi})$ 
by integrating out the non-selected spin degrees of freedom from the full Wigner function in
Eq.~(\ref{theWignerFunction}) [or, equivalently, by replacing $\hat{\rho} \to
\hat{\rho}_I$ in Eq.~(\ref{theWignerFunction})] and then take the equal-angle slice to plot it as
the corresponding correlation function ${\rho}_I$.
For instance, for a system with $N=6$ spins, 
${\rho}_{12345}$ is the equal-angle slice of the reduced Wigner function
\begin{equation}
    W_{12345}(\boldsymbol{\theta},\boldsymbol{\phi}) = \frac{1}{2\pi}\int_0^{\pi}\sin\theta_6 \; \mathrm{d} \theta_6 
    \int_0^{2\pi} \mathrm{d}\phi_6\, W(\boldsymbol{\theta},\boldsymbol{\phi})\,,
    \label{reduced-Wigner}
\end{equation}
such that
\begin{equation}
    \rho_{12345}(\theta,\phi) \equiv W^{\mathrm{EA}}_{12345}(\theta,\phi)\,.
    \label{reduced-rho}
\end{equation}
 Likewise, all of the correlation functions shown in Fig.~\ref{ChainCorrelationKey} 
refer to the equal-angle slices of the 
corresponding reduced Wigner functions.

\section{\label{SpinChains}Applications to Spin Systems}

\begin{figure}
    \includegraphics[scale=0.6]{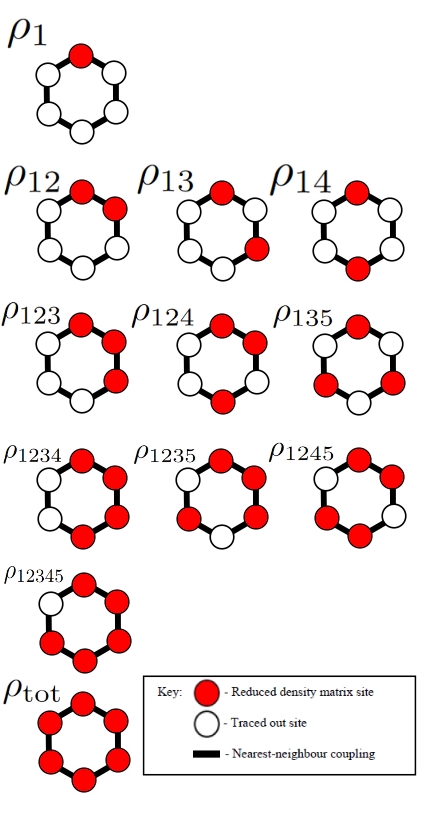}
    \caption{\label{ChainCorrelationKey} Key showing correlation functions 
    explored in each model. The symmetry of the system and the fact that the 
    particles are identical enable us to reduce the number of correlation 
    functions needed to explore the systems completely.}
\end{figure}

\begin{figure*}
    \includegraphics[width=\textwidth]{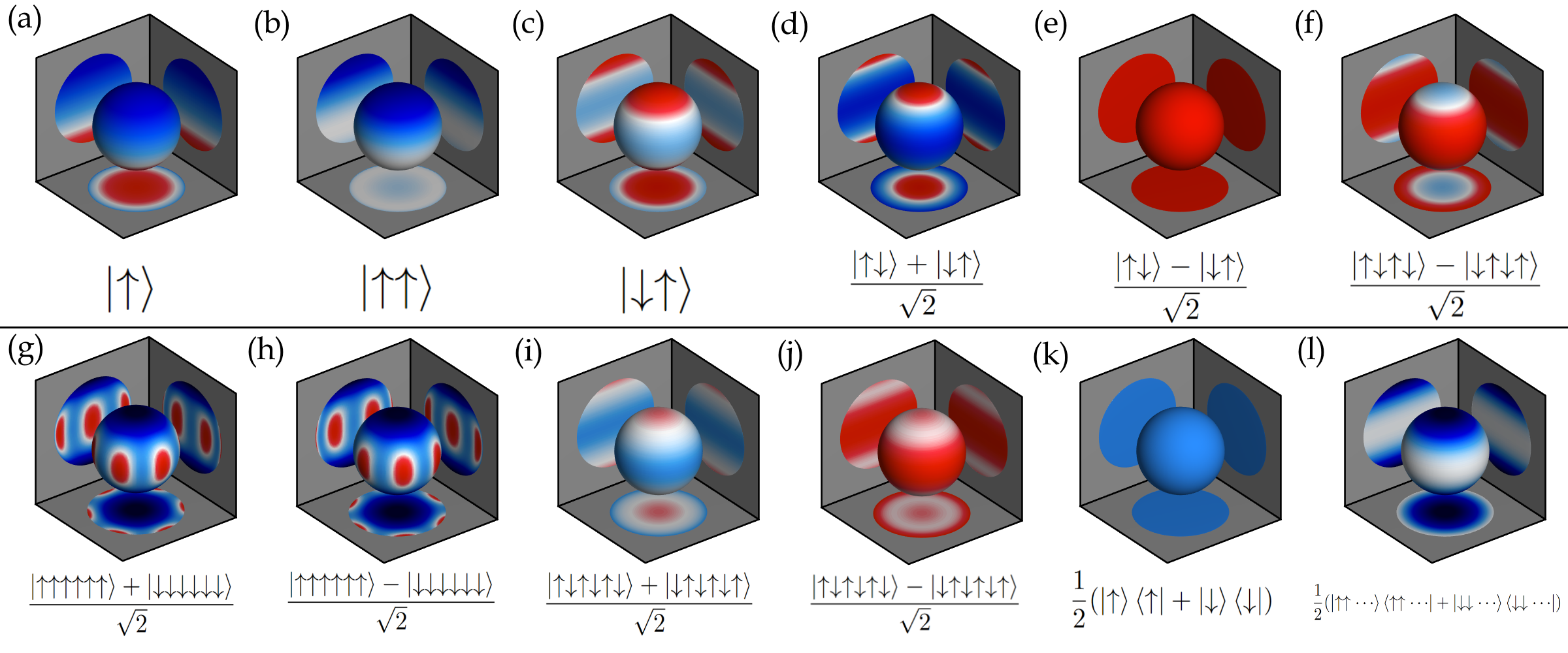}
    \caption{\label{ref_plots} Set of reference spin Wigner functions for states aligned 
    along the $z$ axis that will be used to interpret the results presented 
    below. Each state's Wigner function has been calculated using the 
    equal-angle slice of Eq.~(\ref{WEA}) where $\theta_i=\theta, \phi_i=\phi$ \cite{Wigner7}. The state vector for each 
    Wigner function is given under its respective plot in the qubit basis.
    (a) is the Wigner function for $\ket{\uparrow}$, which is the only state shown [apart from (k)] that doesn't require the equal-angle slice to visualize. 
    The Wigner function for all single-qubit pure states are of a similar form, where they only differ in the placement of the maximum value, that corresponds to the direction on the Bloch sphere. 
    In (b)-(l) we now need to display the functions in the the equal-angle slice. 
    The first of these states is the two-qubit product state $\ket{\uparrow\uparrow}$ in (b). 
    Since this is a doubled product state, each point in the equal-angle slice is equal to the squared value of the equivalent point in the single-qubit phase space, resulting in an equal-angle slice that is non-negative. 
    If instead we want to take a product of the state $\ket{\uparrow}$ with the orthogonal $\ket{\downarrow}$, we get the state in (c).
    The single-qubit Wigner function of $\ket{\downarrow}$ is the same as in (a) with the maximum of the distribution pointing at the south pole, resulting in a Wigner function that is flipped around the equator.
    The equal-angle slice of the product of these two states is then the point-wise multiplication of the individual distributions, resulting in (c).
    Alternatively one may be interested in two-qubit entangled states that start to manifest correlations not possible in product states.
    These can be seen in (d) and (e), that give us an insight into the behavior of entangled states in phase space.
    A comparable four-qubit state to that of (e) is shown in (f).
    More iconic is the appearance of the GHZ states, $\ket{\mathrm{GHZ}_z^\pm}$ in (g) and (h), that manifest in phase space as two coherent states, one on each of the poles, with oscillating positive and negative values around the equator; note that the number of oscillations matches the number of qubits that make up the GHZ state.
    Alternative entangled states are shown in (i) and (j) where entanglement isn't between aligned states; such states will be found later.
    We will also need to consider mixed states as reduced Wigner functions will also be important in our analysis. 
    When taking the reduced Wigner function for one qubit of a multi-qubit maximally entangled state, the result is the fully mixed state in (k) that presents a uniform value over all phase space.
    Further, when we remove the quantum correlations of a maximally entangled GHZ state, the result is a statistical mixture of two spin coherent states where the quantum correlations around the equator are no longer present.
    An example of this can be seen in (l).
    }
\end{figure*}

In this Section we apply the generalized Wigner function formalism 
to a range of physical models: the spin-$\frac{1}{2}$ Ising ferromagnetic cyclic
chain in a transverse magnetic field, the spin-$\frac{1}{2}$ anisotropic $XY$ cyclic
chain in a transverse magnetic field, and the spin-$\frac{1}{2}$ $XXZ$ anisotropic
cyclic Heisenberg chain. These models have been chosen due to the
wide range of quantum phase transitions that they exhibit \cite{SpinChain1B}, 
which in turn enables us to demonstrate the utility of phase-space 
techniques in exploring quantum critical behavior. Each model is
considered as a 1D chain of identical spin-$\frac{1}{2}$ particles, 
with periodic boundary conditions and a finite number of spins $(N=6)$. 
The finite number of spins  enables exact solutions of the ground state to be calculated 
through direct diagonalization of the respective Hamiltonian
of each model \cite{SpinChain1B}. 
The finite size also keeps the dimensionality of the 
system relatively low to enable more intuitive analysis of the results. 
In addition to this, the low number of spins and the GWFs ability to 
visualize high dimensional systems enables an exhaustive exploration of 
single, bipartite and multipartite correlations within the models, 
which are known to provide valuable insights into critical behavior 
\cite{QuantCorrelations1,QuantCorrelations4}. 

Figure~\ref{ChainCorrelationKey} 
shows all correlations explored in these models. Previous demonstrations 
of critical behavior in small finite-sized
systems have demonstrated their utility in exploring multi-body quantum systems 
and QPTs \cite{TI_QPT_1, XY_QPT_1}.  Explicit discontinuities in the 
phase lines at critical points and more exotic phases are not always 
expected however, as our models are not implemented in the thermodynamic 
limit $(N\rightarrow \infty)$ \cite{QuantPhaseTrans,QuantPhaseTrans2}. Due to this 
limitation we have chosen to include spin Wigner functions plotted on the 
Bloch sphere for each correlation function of the system to better 
witness and characterize critical behavior by enabling a deeper 
appreciation of the state of the system at points of interest. This 
also enables us to infer the state of the system and its sub-systems 
at these points.  As we show in detail later, we can thereby infer the presence of critical behavior 
 in an infinite 
system. Figure~\ref{ref_plots} shows a collection of reference 
spin Wigner functions that will be used to interpret the results 
presented below.

\subsection{\label{TI}Spin-1/2 Transverse-Field Ising Spin Chain Model}

The Hamiltonian for a spin-$\frac{1}{2}$ 1D transverse-field Ising model with 
periodic boundary conditions is given by
\begin{equation}
    \hat{\mathcal{H}}=-\sum_{i=1}^{N}\left[\lambda\hat{\sigma}^x_i \hat{\sigma}^x_{i+1} + h\hat{\sigma}^z_i\right]\,,
    \label{TI_H}
\end{equation}
where $N$ is the number of spins, $\lambda$ is the strength parameter of the nearest-neighbor Ising
exchange interactions (which are ferromagnetic when $\lambda > 0$, as henceforth assumed
here), $\hat{\sigma}_i^\alpha,\,\alpha = x,y,z$ are the usual Pauli matrices, and $h$ is the external magnetic
field strength  \cite{SpinChain1B,XY_QPT_7,XY_QPT_6,Pfeuty_1970}. 
Periodic boundary conditions are assumed, so that $\hat{\sigma}^{\alpha}_{N+1} = \hat{\sigma}^{\alpha}_{1},\, \alpha =x,y,z$.   
We are at complete liberty to set the overall energy scale, and for simplicity we 
will henceforth set $h=1$. 

Let us first consider the model at the classical level, in which case $\boldsymbol{\sigma}_i$ ($=2\mathbf{s}_i$)
is an arbitrary vector of unit length.  In the presence of the external magnetic field the classical 
spins all cant at an angle $\alpha$ to the $x$ axis, and it is easy to see that the classical 
ground-state energy, $E_0^{\mathrm{cl}}$, is minimized for $\lambda \geq \frac{1}{2}$ by the choice
$\alpha = \sin^{-1}(\frac{1}{2\lambda})$. A classical phase transition then ensues at
$\lambda = \lambda_c^{\mathrm{cl}}=\frac{1}{2}$ such that for $\lambda < \lambda_c^{\mathrm{cl}}$ the
classical spins all align along the $z$ direction of the external field, with $\alpha=\frac{\pi}{2}$.
The ground-state energy per spin is thus given at the classical level by
\begin{equation}
    \frac{E_0^{\mathrm{cl}}}{N} =
    \begin{cases}
    -\frac{1+4\lambda^2}{4\lambda}\,; \,&\lambda \geq \frac{1}{2}\\
    -1\,;\,&\lambda<\frac{1}{2}\,.
    \end{cases}
\label{TI-energy-cl}
\end{equation}
Clearly, both the energy and its first derivative with respect to $\lambda$ are continuous
at the classical transition point.  Similarly, the classical values of the components of 
the magnetization (viz., the particle spin) vector in both the $x$ (viz., the Ising) 
and the $z$ (viz., the transverse field) directions are given by
\begin{equation}
    M_{\mathrm{cl}}^x =
    \begin{cases}
    \frac{1}{2}\sqrt{1-\frac{1}{4\lambda^2}}\,;\,&\lambda \geq \frac{1}{2}\\
    0\,;\,&\lambda<\frac{1}{2}\,,
    \end{cases}
\label{TI-magx-cl}
\end{equation}
and
\begin{equation}
    M_{\mathrm{cl}}^z =
    \begin{cases}
    \frac{1}{4\lambda}\,;\,&\lambda \geq \frac{1}{2}\\
    \frac{1}{2}\,;\,&\lambda<\frac{1}{2}\,.
    \end{cases}
\label{TI-magz-cl}
\end{equation}

The quantum spin-$\frac{1}{2}$ version of the model is also exactly solvable for both finite values of $N$ 
and in the thermodynamic limit ($N \to \infty $) through the following sequence
of transformations: (i) a Jordan-Wigner transformation~\cite{Jordan-Wigner_1928}, 
which transforms the spin operators into
fermionic operators; (ii) a Fourier transformation from lattice position to
lattice momentum; and (iii) a Bogoliubov transformation \cite{SpinChain1B,Pfeuty_1970}. It is
important to note too that the model possesses a ${\mathbb Z}_2$ symmetry group due 
to its invariance under the unitary operation of flipping all spins in the $x$ direction. 
It is precisely the breaking of this symmetry that causes the sole QPT that this model
possesses in the thermodynamic limit ($N \to \infty$) \cite{Pfeuty_1970}.  The exact
ground-state energy per spin for the quantum spin-$\frac{1}{2}$ model of
Eq.~(\ref{TI_H}) in the thermodynamic limit ($N\to\infty$) is given by \cite{Pfeuty_1970}
\begin{equation}
    \frac{E_0}{N}=-\frac{1}{\pi}\int_{0}^{\pi}\mathrm{d}k\sqrt{1+2\lambda\cos k+\lambda^2}\,,
\label{TI-energy}
\end{equation}
which is an elliptic integral of the second kind.
Although the expressions of Eqs.~(\ref{TI-energy-cl}) and (\ref{TI-energy}) are very
dissimilar, they agree numerically with one another to a few percent or less.  For example,
at $\lambda=1$, we have $E_0^{\mathrm{cl}}(\lambda=1)/N=-1.25$ and 
$E_{0}(\lambda=1)/N = -\frac{4}{\pi}\approx -1.273$.  It is also trivial to confirm that in
the two limiting cases $\lambda\to\infty$ and $\lambda\to 0$, the expression of 
Eq.~(\ref{TI-energy}) reduces respectively to the values $E_{0}(\lambda\to\infty)/N \to -\lambda$
and $E_{0}(\lambda=0)=-1$, which are exactly also as given by the classical expression of
Eq.~(\ref{TI-energy-cl}).  The agreement of the classical and quantum results for the
ground-state energy in these two limiting cases is exactly as expected, since in both of these two
extremes the spins are fully aligned, and such fully aligned states are also eigenstates of
the corresponding quantum Hamiltonian.

Despite the above levels of numerical agreement, the classical and quantum results for the 
ground-state energy differ in one profound aspect.  Thus, the integral expression
in Eq.~(\ref{TI-energy}) is nonanalytic at the point $\lambda=\lambda_c =1$,
which is now the quantum phase transition point, which is considerably shifted from its
classical counterpart at the quite different value $\lambda = \lambda_c^{\mathrm{cl}}=\frac{1}{2}$.
One may readily confirm from Eq.~(\ref{TI-energy}) that both $E_0$ and $\partial E_0/\partial\lambda$
are continuous and finite at $\lambda=\lambda_c =1$, whereas all higher-order derivatives
$\partial^{n}\!E_0/\partial\lambda^n$ with $n>1$ diverge at the same point, so that 
the transition there is of 2QPT type.

The exact result for the $x$ component (viz., in the Ising direction) of the magnetization (i.e.,
the ground-state expectation value of the spin vector, $\mathbf{s}_i=\frac{1}{2}\boldsymbol{\sigma}_i$,
at a given site) for the system in the thermodynamic limit ($N\to\infty$) is given by \cite{Pfeuty_1970}
\begin{equation}
    M^x=
    \begin{cases}
    \frac{1}{2}(1-\frac{1}{\lambda^2})^{1/8}\,;\,&\lambda \geq 1\\
    0\,;\,&\lambda<1\,,
    \end{cases}
\label{TI-magx}
\end{equation}
which exhibits the quantum phase transition at $\lambda=\lambda_c =1$ much more clearly
than does Eq.~(\ref{TI-energy}) for the ground-state energy.  Finally, the 
corresponding exact result for the $z$ component (viz., in the transverse field direction) of
the magnetization for the system in the thermodynamic limit ($N\to\infty$) is given by \cite{Pfeuty_1970}
\begin{equation}
     M^z=\frac{1}{2\pi}\int_{0}^{\pi}\mathrm{d}k\,\frac{(1+\lambda\cos k)}{\sqrt{1+2\lambda\cos k +\lambda^2}}\,,
\label{TI-magz}    
\end{equation}
an elliptic integral of the first kind.
Once again, Eq.~(\ref{TI-magz}) is nonanalytic at the quantum phase transition point, $\lambda_c =1$,
where it may also readily be evaluated, $M^{z}(\lambda=1)=\frac{1}{\pi}$. One may readily confirm
from Eq.~(\ref{TI-magz}) that as $\lambda$ increases from zero to infinity $M^z$ decreases smoothly
and monotonically from $\frac{1}{2}$ to zero, the same extreme values as in the corresponding
classical result of Eq.~(\ref{TI-magz-cl}).

Thus, in the limit $\lambda \to \infty$, the model has perfect ferromagnetic order
with all the spins aligned along either the positive or negative $x$ direction.  This two-fold
degeneracy of the ground state is maintained  for the system in the thermodynamic 
limit as $\lambda$ is decreased (and, accordingly, the magnetization in the spin-space 
$x$ direction decreases from its maximal value at $\lambda \to \infty$)  until at the
critical point $\lambda_{c}=1$ the magnetization in the $x$ direction disappears.  For
all values $\lambda > \lambda_c$ the ground state of the system is a (ferromagnetic) ordered state
in which the spin-spin interactions are able to prevail over the interaction with the 
external magnetic field, and the system thus has a non-vanishing value for
the magnetization in the $x$ direction (which now plays the role of order parameter).  By
contrast, for $0 < \lambda < \lambda_c$ the system is in a disordered state, which is a 
paramagnet (i.e., with vanishing magnetization in the $x$ direction), and which
preserves the ${\mathbb Z}_2$ symmetry of the Hamiltonian, and is hence non-degenerate.  For
example, in the limiting case $\lambda = 0$ the ground state has all of the spins aligned 
in the $z$ direction (of the external field). Both the ferromagnetic and paramagnetic phases 
are gapped.  It is only at the precise 2QPT point for the infinite ($N \to \infty$) chain, 
$\lambda = \lambda_c =1$, that the ground state is gapless.

We note that the presence of a ground-state degeneracy in general  always implies the need 
to choose among the possible ground states.  For the sake of consistency between regions with
degeneracy (e.g., as in the ordered ferromagnetic phase of the current Ising model) and without
degeneracy (e.g., as in the disordered paramagnetic phase of the current Ising model), we shall
always choose the ground state to preserve all of the symmetries of the Hamiltonian.  Thus, for
the case of double degeneracy, as here, this choice simply implies a ground state, 
in the limiting case $\lambda\to\infty$ of the pure Ising model, for example, 
of the Greenberger-Horne-Zeilinger (GHZ) type \cite{GHZ_2007},
both in the thermodynamic limit and for finite values of $N$,
which we discuss in more detail below. 

Consider the case of $N$ finite first. Thus for zero external field (i.e.,
$\lambda \to \infty$) the ground state is doubly degenerate, with the two fully
aligned states having values $M^x = \pm\frac{1}{2}$ of the magnetization.
This degeneracy is then lifted by the addition of the external magnetic 
field, as in Eq.~(\ref{TI_H}) with $\lambda$ finite, so that the new ground state
is symmetric, and is unchanged under the transformation $\sigma_i^x\to -\sigma_i^x$, such that
$M^x =0$, while the first excited state is antisymmetric.  It has been shown \cite{Pfeuty_1970}
that these two states have a gap of order $\lambda^{-N}$ for $\lambda>1$, which hence disappears
rapidly as $N\to\infty$.  Thus, in the thermodynamic limit ($N\to\infty$), the ground 
and first excited states become degenerate, allowing a non-zero order parameter.  Conversely, for $\lambda <1$, the ground
state remains non-degenerate in the thermodynamic limit, and no order appears (i.e., $M^x =0$).

As stated previously, our main aim here is to demonstrate the utility of our generalized
Wigner function formalism to detect, characterize and distinguish QPTs of various kinds,
including the 2QPT that is exhibited by the present model, as discussed above. 
The finite size ($N=6$) that we choose
for our calculations of the system below enables exact solutions to be calculated through direct 
diagonalization of its Hamiltonian and the subsequent evaluation of the density matrix 
and the complete set of reduced density matrices (as shown in Fig.~\ref{ChainCorrelationKey}).

\begin{figure}
    \includegraphics[width=\linewidth]{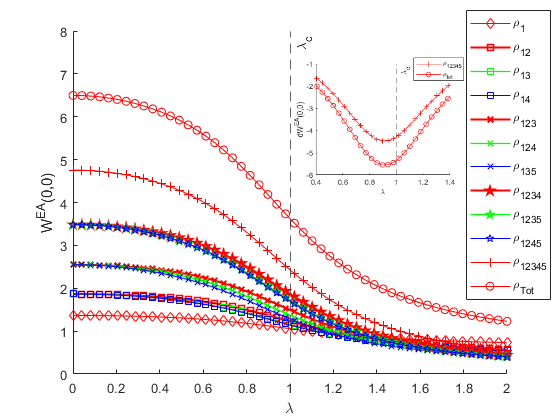}
    \caption{\label{TIPhaseLine} Equal-angle slices of the reduced GWFs,
    $\rho_{I}(\theta,\phi) \equiv W^{\mathrm{EA}}_{I}(\theta,\phi)$, 
    shown in Fig.~\ref{ChainCorrelationKey}, taken at $\theta=\phi=0$, for 
    the $N=6$ spin-$\frac{1}{2}$ Ising ferromagnetic cyclic
    chain in a transverse magnetic field [see Eq.~(\ref{TI_H})] with $h=1$. 
    The 2QPT critical point at $\lambda_{c}$ is marked on the plot
    The insert shows the first derivatives with respect to $\lambda$ of 
    ${\rho}_{12345}$ and ${\rho}_{\mathrm{tot}}$, thereby highlighting 
    the critical point of the system to be $\lambda_{c}\approx0.9$. 
    A full sized version of this plot is available in the supplementary material.}
\end{figure}
    
\begin{figure*}
    \includegraphics[width=\textwidth]{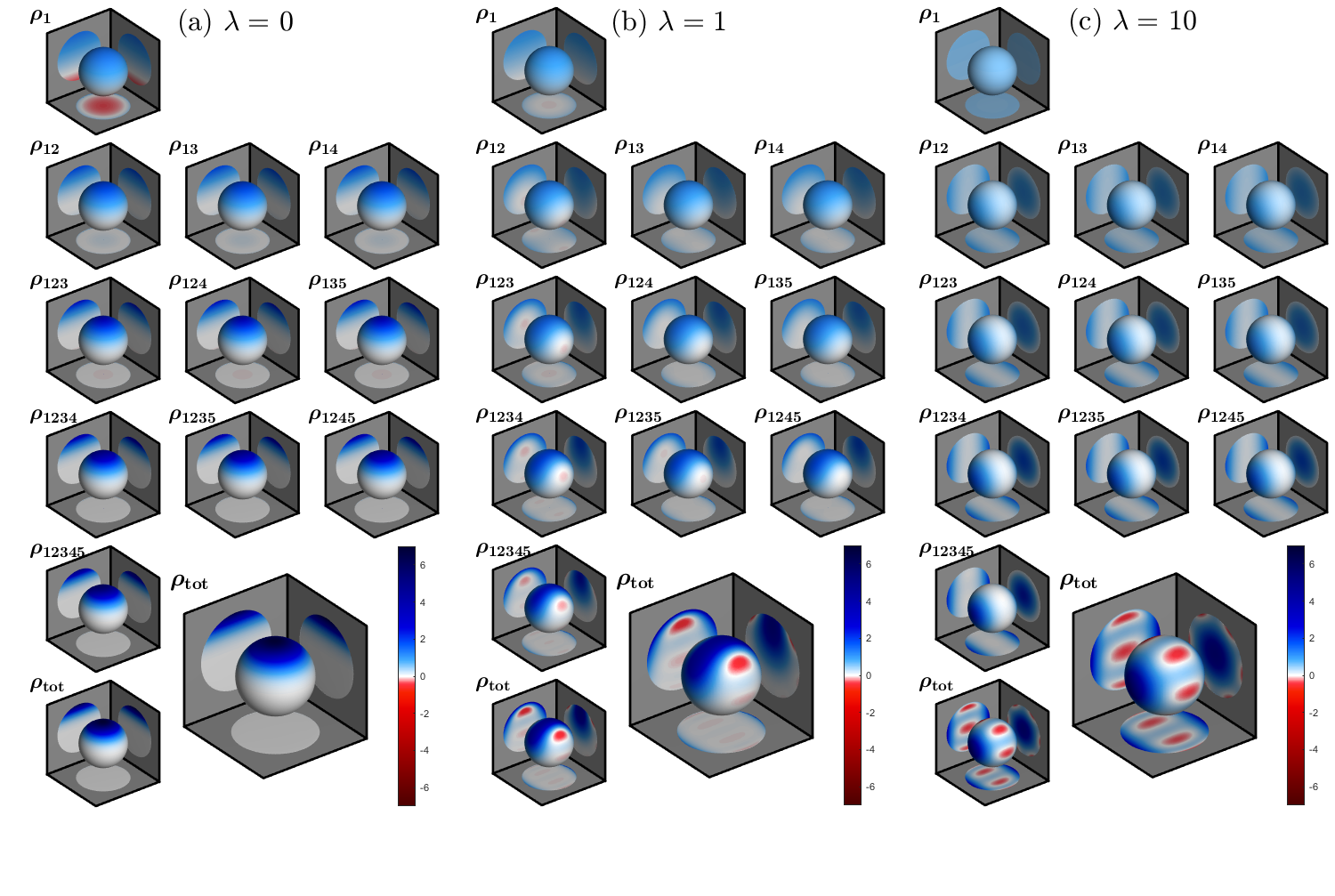}
    \caption{\label{TISpinPlots} Equal-angle slices of the reduced GWFs,
    $\rho_{I}(\theta,\phi) \equiv W^{\mathrm{EA}}_{I}(\theta,\phi)$, 
    shown in Fig.~\ref{ChainCorrelationKey}, plotted on the Bloch sphere.
    Results are shown for the $N=6$ spin-$\frac{1}{2}$ Ising ferromagnetic cyclic
    chain in a transverse magnetic field [see Eq.~(\ref{TI_H})], 
    with $h=1$, at three points of interest; (a) $\lambda=0$, (b) $\lambda=\lambda_{c}=1$, and (c) $\lambda=10$.
    ${\rho}_\mathrm{tot}$ is replicated in larger subplots to enable a better appreciation 
    of the state of the system, including any intricacies that arise from multi-qubit Wigner functions. 
    An animation showing all spin Wigner plots across the QPT is contained in the supplementary material.
    The initial value of $\lambda = 0$ in (a) eliminates any $\hat{\sigma}_x$ terms from the Hamiltonian. 
    This results in a ground state that is fully aligned along
    the  $z$  axis, where the state is fully separable and each spin is in the same state.
    (b) shows the ground state for $\lambda=1$; in the infinite chain case this is a critical point of the 2QPT. 
    In the finite-chain case here, we see signs of the disordered paramagnetic phase; in phase space, 
    we see this as a gradual shift towards a GHZ state aligned along the $x$ axis.
    This trend is then continued in (c), and as $\lambda$ increases further the ground state 
    finally reaches the state $\ket{\mathrm{GHZ}_x^+}$ at $\lambda\to\infty$.
    }
\end{figure*}

Figure~\ref{TIPhaseLine} shows the behavior of the $N=6$ transverse Ising model 
as the coupling parameter $\lambda$ is varied in terms of equal-angle slices 
of the GWF for the correlation functions shown in Fig.~\ref{ChainCorrelationKey}. 
We choose to take the equal-angle slice of the GWF 
$(\theta_{1}=\theta_{2}=\theta_{3}=\theta_{4}=\theta_{4}=
\theta_{5}=\theta_{6}\equiv\theta$,
$\phi_{1}=\phi_{2}=\phi_{3}=\phi_{4}=\phi_{5}=\phi_{6}\equiv\phi)$, with 
$\theta=\phi=0$, as this enables appreciation of high-dimensional systems while also 
remaining invariant under particle exchange \cite{Wigner5}. 
Clear discontinuities are not seen in the phase line plot of Fig.~\ref{TIPhaseLine}  
for the transverse Ising model due both to the finite system size and the 
nature of 2QPTs. A continuous gradual change in the value of the GWF for each correlation function 
is seen across the range with higher-order correlations experiencing a larger shift. 

It has been shown elsewhere that exploration of the first derivative of the GWF can witness 
and characterize 2QPT critical points~\cite{Discrete}. Accordingly, we plot the first derivatives 
with respect to $\lambda$ of ${\rho}_{\mathrm{tot}}$ and ${\rho}_{12345}$ in 
the insert to Fig.~\ref{TIPhaseLine}. Both derivatives experience 
a minimum at $\lambda \approx 0.9$, highlighting the presence of the 2QPT pseudo-critical 
point for the finite system. We  note that all of the other correlation functions 
shown in Fig.~\ref{TIPhaseLine} experience a very similar minimum in their first derivatives. 
The slight deviation from the minimum away from the infinite-chain critical point $\lambda_{c}=1$ is 
almost certainly due to the finite size of the system \cite{XY_QPT_2,XY_QPT_4}. 
A similar finite-size scaling behavior is also witnessed in the transverse-field 
$XY$ model discussed below in Sec.~\ref{XY}. 
Interestingly however, the pseudo-critical point of the current transverse-field Ising
model seems to occur at decreasing values as the size of the system decreases, rather than at the
increasing values seen in the corresponding transverse-field $XY$ model. 
This scaling behavior is worthy of further exploration.

Figure~\ref{TISpinPlots} shows the spin Wigner plots at points of interest in the 
transverse-field Ising model. Figure~\ref{TISpinPlots}(a) shows the state of 
the system at the initial value $\lambda=0$. All correlation functions in the system clearly correspond to a
state with all spins aligned along the $z$ axis [and see Figs.~\ref{ref_plots}(a) and (b)] 
at this point. 

Figure~\ref{TISpinPlots}(b) shows the state of the system at the critical point 
$\lambda_{c} = 1$ of the 2QPT in the infinite chain, at which point the finite chain
should be showing signs of the disordered paramagnetic phase in the infinite chain,
according to the results shown in the insert to Fig.~\ref{TIPhaseLine}, and as discussed above. Thus, the 
GWF plot for ${\rho}_{\mathrm{tot}}$ in Fig.~\ref{TISpinPlots}(b) clearly witnesses an $x$-axis aligned 
GHZ type state [see Fig.~\ref{ref_plots}(g) for the corresponding GHZ state in
the $z$ direction], with all of the lower-order correlations also presenting features of 
the GHZ state. 

Finally, Fig.~\ref{TISpinPlots}(c) shows the state of the system at $\lambda=10$. 
The GWF plot for ${\rho}_{\mathrm{tot}}$ now clearly shows a more uniform $x$-axis aligned  GHZ state,
with the south-pole region at a higher amplitude than for the previous case $\lambda = 1$. 
${\rho}_{12345}$ continues to present features of the GHZ state now with lower amplitude nodes of 
negativity. Lower order correlations no longer present clear features of the GHZ state with many 
tending towards a mixed state. 

The Hamiltonian of Eq.~(\ref{TI_H}) is stoquastic (i.e., all off-diagonal matrix elements are non-positive) in the basis
$\{\ket{\rightarrow}=(\ket{\uparrow}+\ket{\downarrow})/\sqrt 2,\,\ket{\leftarrow}=(\ket{\uparrow}-\ket{\downarrow})/\sqrt 2\}$ 
for $\lambda\ge0$; therefore the ground state can be written as a positive linear combination 
in this basis.  For finite $N$ the ground state evolves from the $z$-axis-aligned state at $\lambda = 0$,
\begin{equation}
    \ket{\uparrow}^{\otimes N}=2^{-\frac{N}{2}}\sum_{s_1=\leftarrow}^{\rightarrow}\sum_{s_2=\leftarrow}^{\rightarrow}\cdots\sum_{s_N=\leftarrow}^{\rightarrow}\ket{s_1s_2\cdots s_N}\,, 
\label{lambda=0}
\end{equation}
to the GHZ state that is a positive superposition of the two $x$-axis Ising-ordered states,
\begin{equation}
    \ket{\mathrm{GHZ}_x^+}=\frac{1}{\sqrt{2}}\left(\ket{\rightarrow}^{\otimes N}+\ket{\leftarrow}^{\otimes N}\right)\,, 
\label{h=0}
\end{equation}
in the limit $\lambda \to \infty$. So long as we take the limit $\lambda \rightarrow \infty$ before the 
thermodynamic limit $N \rightarrow \infty$, an alternative argument is that as the coupling is increased
adiabatically from the value $\lambda=0$, where it contains the equally weighted admixture of the $2^N$ terms indicated 
in Eq.~(\ref{lambda=0}), all but the two fully aligned terms of Eq.~(\ref{h=0}) will smoothly disappear as
$\lambda\to\infty$ and, by symmetry, their phase relationship will be retained.
The entanglement therefore increases with $\lambda$; in the GHZ limit given by Eq.~(\ref{h=0}) all reduced 
density matrices are mixtures $\frac{1}{2}(\ket{\rightarrow\cdots\rightarrow}\bra{\rightarrow\cdots\rightarrow
}\,+\,\ket{\leftarrow\cdots\leftarrow}\bra{\leftarrow\cdots\leftarrow})$.  Accordingly, only the full Wigner 
function will show interference terms, as seen in Fig 4(c). 

\subsection{\label{XY} {Spin-1/2 Transverse-Field \it XY} Spin Chain Model}

The spin-$\frac{1}{2}$ transverse-field  anisotropic 
$XY$ model on a 1D chain with periodic boundary conditions is a generalization of its Ising 
counterpart that we have discussed above in Sec.~\ref{TI}.  Its Hamiltonian is given by
\begin{equation}
\hat{\mathcal{H}} = -\sum_{i=1}^{N}\left\{\frac{\lambda}{2}[(1+\gamma)\hat{\sigma}^x_i \hat{\sigma}^x_{i+1} + 
(1-\gamma)\hat{\sigma}^y_i \hat{\sigma}^y_{i+1}]+h\hat{\sigma}_i^z\right\} \,,
\label{XY_H}
\end{equation}
where $\gamma$ is now the spin anisotropy parameter, and the remaining parameters
are exactly as in the Hamiltonian of Eq.~(\ref{TI_H}) for its corresponding
Ising limit counterpart, and to  which Eq.~(\ref{XY_H}) reduces for the special
case $\gamma = 1$.  Once again, periodic boundary conditions are assumed, such that
$\hat{\sigma}^{\alpha}_{N+1} = \hat{\sigma}^{\alpha}_{1},\, \alpha =x,y,z$.   Just 
as in Sec.~\ref{TI} for the special case $\gamma = 1$ of the transverse-field Ising
model, we again set the external field strength to $h=1$ in order to set the overall energy
scale and to simplify the parameter set.  We also restrict ourselves to the case where
the nearest-neighbor spin interactions are wholly ferromagnetic, such that $\lambda > 0$
and $-1 \leq \gamma \leq 1$.  (Clearly, it is actually sufficient to restrict ourselves
to the range $0 \leq \gamma \leq 1$, since the Hamiltonian is invariant under the replacements
$\hat{\sigma}_i^x \leftrightarrow \hat{\sigma}_i^y, i=1,\dots, N;\, \gamma \rightarrow -\gamma$).
The special case $\gamma=0$ is simply the transverse-field isotropic $XY$ (i.e., the 
transverse-field $XX$) model.

The Hamiltonian of Eq.~(\ref{XY_H}) again has a ${\mathbb Z}_2$ symmetry associated 
with the fact that it is invariant under spin rotations of $\pi$ about the global spin-space
$z$ axis, under which $\hat{\sigma}_i^\mu \rightarrow -\hat{\sigma}_i^\mu\,, \mu = x,y\,; i=1,\dots,N$. 
It thus commutes with the spin parity operator defined as
\begin{equation}
    \hat{P}_z \equiv (-i)^{N} \exp (i\pi\sum_{l=1}^N \frac{\hat{\sigma}_l^z}{2}) = \prod_{l=1}^N \hat{\sigma}_l^z\,.
\label{spin-parity-op}    
\end{equation}
Hence, all {\it non-degenerate} eigenstates of the Hamiltonian of Eq.~(\ref{XY_H}), including the 
ground state, are also eigenstates of $\hat{P}_z$ with eigenvalues $\pm 1$. At the $XX$ isotropic limiting 
point ($\gamma = 0$) of the model, the Hamiltonian has a much larger symmetry group, since it
is now invariant under rotations by an {\it arbitrary} angle about the global spin-space
$z$ axis.

The transverse-field spin-$\frac{1}{2}$ anisotropic $XY$ model 
on the 1D  chain with periodic boundary conditions
described by Eq.~(\ref{XY_H}) plays an important archetypal role in quantum many-body theory
and quantum statistical mechanics for two important reasons. First, it is one of the 
relatively rare models for which an exact analytic solution is known, as described below. 
Second, the model provides a good approximation to a variety of real physical systems,
which can hence be used to simulate it.  Examples include ultra-cold neutral atoms loaded onto
an optical lattice~\cite{Duan-Demler-Lukin_2003}, and a quantum circuit that processes 
$\log N$ qubits, which simulates the model with $N$ spins (or qubits)~\cite{Boyajian-et-al_2013}.

We note that the model, like its special-case ($\gamma = 1$) Ising counterpart, is also
exactly solvable both for all finite values of $N$ and in the thermodynamic limit ($N \to \infty$) by
exactly the same sequence of transformations as discussed in Sec.~\ref{TI}. In the case of zero external 
field the solution was first obtained by Lieb, Schultz, and Mattis~\cite{XY_Lieb-et-al_1961}. This
method of solution is easily generalized to the case of nonzero external field for any finite
value of $N$ (see, e.g., Ref.~\cite{XY_Rossignoli-et-al_2008}), and the exact 
solution in the thermodynamic limit ($N \to \infty$) has also been discussed
separately~\cite{StatHXY, StatHXY2}. Of course, for the small-chain ($N=6$) results presented
here, direct diagonalization of the Hamiltonian is readily performed, and the ground-state
density matrix and subsequent reduced matrices of the system can then be calculated,

In the $\lambda$--$\gamma$ plane, in our region of interest (viz., when $h=1,\, \lambda >0,\,
|\gamma|<1$), the phase diagram of the system in the thermodynamic limit is separated into
three regions by the lines $\lambda =1$ and $\{\gamma=0,\, \lambda \geq 1\}$.
The line segment $\{\gamma=0,\, \lambda \geq 1\}$ simply demarcates the anisotropic phase
transition from a ferromagnetic phase with magnetic ordering in the spin-space $x$ direction
(for $\gamma >0$) to one ordered in the spin-space $y$ direction (for $\gamma <0$).
The line $\lambda = \lambda_c \equiv 1$ denotes a line of Ising 2QPTs from the ordered
ferromagnetic phases (for $\lambda >1$) to the disordered (i.e., with
vanishing magnetization in the spin-space $x$--$y$ plane) paramagnetic phase 
(for $0<\lambda<1$) \cite{XY_QPT_1,XY_QPT_2,XY_Patra-et-al_2011,XY_QPT_3,XY_QPT_4}.

In the latter paramagnetic phase (for $\lambda<\lambda_c$) the ground state is non-degenerate,
and the energy spectrum shows a gap to the first excited state for all values of $\gamma$,
By contrast, in the ordered phases (for $\lambda>\lambda_c$) the system has a doubly
degenerate ground state with a gapped energy spectrum for all nonzero values of $\gamma$,
while precisely at the isotropic point $\gamma =0$ the ground state is non-degenerate
and the energy spectrum is gapless.  The spectrum is again gapless precisely on
the critical line $\lambda = \lambda_c$. Clearly, the point $(\lambda=1,\,\gamma =0)$ is
a multicritical point in the quantum phase diagram of the model.  We note that it has also
been shown~\cite{XY_QPT_2,XY_QPT_4} that finite-sized chains display a pseudo-critical
point $\lambda_c(N)$, which deviates from the true critical point $\lambda_c = \lambda_c(\infty)$,
such that $\lambda_c(N)>\lambda_c$ for all finite values of $N$, and 
$\lambda_c(N) \to \lambda_c$ as $N \to \infty$.  With no loss of generality, as 
explained above, we henceforth restrict ourselves to values of the anisotropy parameter
in the range $0 \leq \gamma \leq 1$.

Once again, as in Sec.~\ref{TI}, we note that, in order to be consistent between regions with
degeneracy (e.g., as in the ordered ferromagnetic phases of the current anisotropic $XY$ 
model for $\gamma \neq 0$) and without degeneracy (e.g., as in the disordered paramagnetic 
phase of the current anisotropic $XY$ model), for all our calculations below we 
always choose the ground state to preserve all of the symmetries of the Hamiltonian.

The Hamiltonian of Eq.~(\ref{XY_H}) also possesses another remarkable property. Namely, for any
value of the anisotropy parameter in the range $0 < \gamma \leq 1$ 
and for all values of the system size $N$, there exists
a ground-state factorization point $\lambda_{f}=\lambda_{f}(\gamma)$, given by
\begin{equation}
\lambda_{f} = \frac{1}{\sqrt{1-\gamma^{2}}} \,,
\label{factorisation}
\end{equation}
for which the system contains two degenerate ground states, both of which are fully
factorized in the sense that they are simple products of single-site states \cite{XY_Rossignoli-et-al_2008,Giampaolo-et-al_2008,XY_Ciliberti-et-al_2010,XY_QPT_3,XY_QPT_1}. 
Clearly, for these states themselves entanglement vanishes in principle.  However, one 
must take note that these factorized states themselves are {\it not} eigenstates of the 
spin parity operator $\hat{P}_z$, as we now discuss more fully.

For the case $0 < \gamma \leq 1$, one can rather readily show~\cite{XY_Rossignoli-et-al_2008} 
that one of the two factorized states, which we denote by $\ket{\vartheta}$, has all of the 
spins fully aligned in the $x$--$z$ spin-space plane and at at an angle $\vartheta$ to the spin $z$ axis,
given by
\begin{equation}
    \vartheta = \cos^{-1}\sqrt{\frac{1-\gamma}{1+\gamma}}\,.
    \label{fac-angle}
\end{equation}
The other factorized state, degenerate in energy with $\ket{\vartheta}$, is simply 
$\ket{-\vartheta}=\hat{P}_z\ket{\vartheta}$.  The two states $\ket{\pm \vartheta}$ are non-orthogonal
for $\vartheta \neq \frac{\pi}{2}$ (i.e., for $\gamma \neq 1$).  In that case the 
correct orthonormal basis that conserves spin parity is spanned by the two {\it entangled \,}
states $\ket{\vartheta_\pm}$, defined by
\begin{equation}
    \ket{\vartheta_\pm} \equiv \frac{\ket\vartheta \pm \ket{-\vartheta}}{\sqrt{2(1\pm \braket{-\vartheta}{\vartheta})}}\,,
    \label{fac-states-with-def-parity}
\end{equation}
which satisfy $\hat{P}_z\ket{\vartheta_\pm} = \pm\ket{\vartheta_\pm}$.
We note that an equal-angle slice of the reduced GWF,
$\rho_{I}(\theta,\phi) \equiv W^{\mathrm{EA}}_{I}(\theta,\phi)$,
for a similar wave function to that of Eq.~(\ref{fac-states-with-def-parity}),
viz., one with $\vartheta=\frac{\pi}{4}$, such that the two aligned states have a relative
angle $\frac{\pi}{2}$ between their alignment directions, is shown in Fig.~6(f) of
Ref.~\cite{Wigner7}. 

The states $\ket{\vartheta_\pm}$, as defined in Eq.~(\ref{fac-states-with-def-parity}), are the actual
eigenstates of the Hamiltonian $\hat{\mathcal{H}}$ of Eq.~(\ref{XY_H}) in each spin parity subspace at the
factorization point $\lambda=\lambda_f$, and are just the corresponding limits of the
exact definite-parity ground eigenstates $\ket{\Psi_{\pm}(\lambda)}$ as $\lambda \to \lambda_f$.
Thus, the factorization point $\lambda_f$ represents a crossing point of the
two lowest opposite-parity levels~\cite{XY_Rossignoli-et-al_2008}, and is hence a parity
transition point.  We note that for the limiting value $\gamma = 0$ (i.e., when the Hamiltonian
becomes isotropic), for which $\lambda_f = 1$ the alignment angle $\vartheta = 0$ 
and the ground-state degeneracy vanishes, in accordance with our discussion above. 
Indeed, the state $\ket{0}$ is a trivial eigenstate
of the Hamiltonian of Eq.~(\ref{XY_H}) in this isotropic limit for {\it all} values of
the parameter $\lambda$.  In the opposite limiting case $\gamma =1$, which is just the 
Ising model in a transverse field, the alignment angle $\vartheta = \frac{\pi}{2}$, and
$\lambda_f$ becomes infinite.  In this case the overlap $\braket{-\frac{\pi}{2}}{\frac{\pi}{2}}=0$,
and, as expected, the states $\ket{\frac{\pi}{2}_\pm}$ simply become $x$-axis aligned 
GHZ states, $\ket{\mathrm{GHZ}_x^\pm}$, one of which has been explicitly displayed in 
Eq.~(\ref{h=0}) above for the case of the positive superposition of the two $x$-axis 
Ising-ordered states.  These states are just the counterparts of the $z$-axis aligned 
states, $\ket{\mathrm{GHZ}_z^\pm}$, the equal-angle slices of the reduced GWFs,
$\rho_{I}(\theta,\phi) \equiv W^{\mathrm{EA}}_{I}(\theta,\phi)$,
of which have been shown in Figs.~\ref{ref_plots}(g) and~(h).

Such GHZ states, although globally entangled, display no two-spin entanglement
(for $N>2$).  As an aside here we note that the entanglement properties of the $N=2$ 
version of the model have also been studied in great detail~ \cite{QuantCorrelations6}).  
By contrast, for $0<\vartheta<\frac{\pi}{2}$, the two-spin entanglement
in the states $\ket{\vartheta_\pm}$ depends critically on the nonzero value of the overlap
$\braket{-\vartheta}{\vartheta}$.  Indeed, the two-spin entanglement has been shown to
undergo a change as $\lambda$ is varied across
$\lambda_f$, being zero exactly at the factorization point.  
For this reason this point has also become known as the entanglement transition 
point~\cite{XY_QPT_1}, although its real 
origin lies in the fact that it is a point of accidental spin-parity
symmetry breaking where two branches of eigenstates with different spin
parities cross one another.

We note finally that, for the system in the thermodynamic limit, at {\it all\,} points in
the ordered phase (for which $\lambda>\lambda_c =1$), which includes the factorization point,
the ground state is exactly doubly degenerate, and hence the factorization point cannot be 
expected to be a prominent feature.  Similarly, the (e.g., entanglement) effects
are small in large anisotropic chains, for which the ground states in each of the two
spin-parity sectors are nearly degenerate in energy.  However, as we shall see explicitly below,
the factorization point is quite prominently visible for small finite cyclic chains
(here with $N=6$).  Indeed, they have also been shown to remain appreciable for 
increasingly larger values of $N$ as the size of the anisotropy parameter $\gamma$
is accordingly decreased~\cite{XY_Rossignoli-et-al_2008}.

\begin{figure}
    \includegraphics[width=\linewidth]{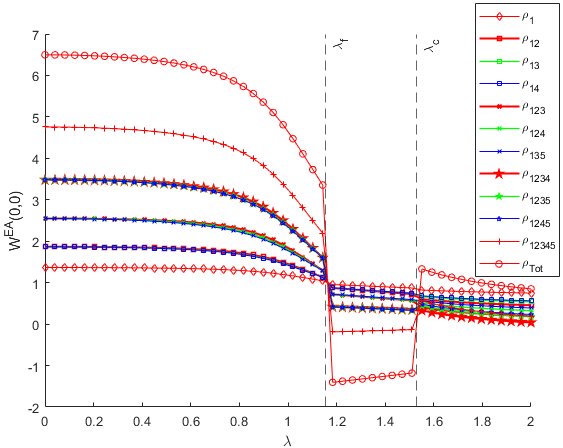}
    \caption{\label{XYPhaseLine} Equal-angle slices of the reduced GWFs,
    $\rho_{I}(\theta,\phi) \equiv W^{\mathrm{EA}}_{I}(\theta,\phi)$, 
    shown in Fig.~\ref{ChainCorrelationKey}, taken at $\theta=\phi=0$, 
    for the $N=6$ spin-$\frac{1}{2}$ anisotropic $XY$ cyclic
    chain in a transverse magnetic field, $h=1$ [see Eq.~(\ref{XY_H})] with 
    $\gamma=0.5$. The factorization point $\lambda_{f}$ and pseudo-critical 
    point $\lambda_{c}$ are marked on the plot. A full-sized version 
    of this plot is available in the supplementary material.
    }
\end{figure}

\begin{figure*}
    \includegraphics[width=\textwidth]{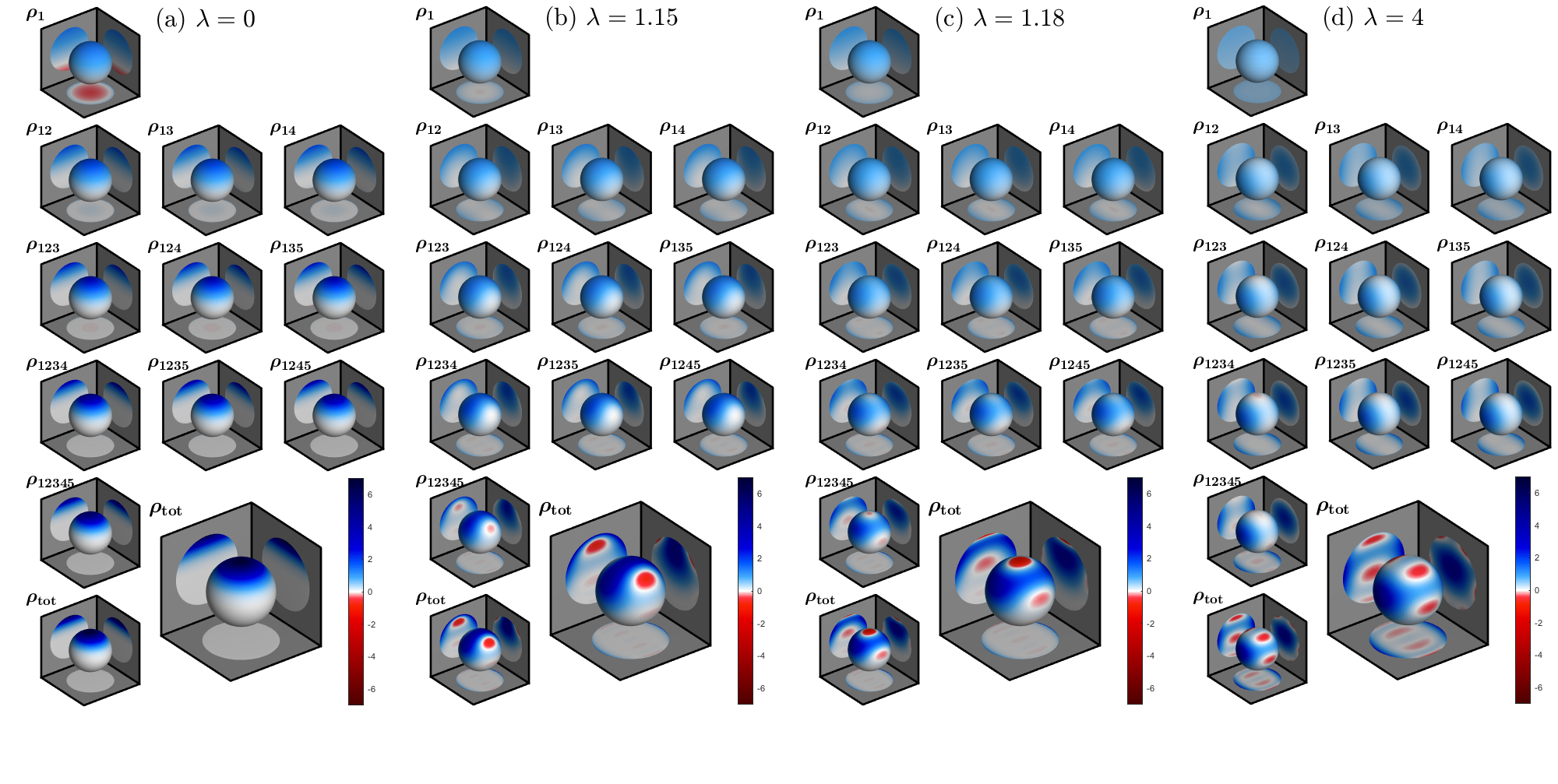}
    \caption{\label{XYSpinPlots} Equal-angle slices of the reduced GWFs,
    $\rho_{I}(\theta,\phi) \equiv W^{\mathrm{EA}}_{I}(\theta,\phi)$, 
    shown in Fig.~\ref{ChainCorrelationKey}, plotted on the Bloch sphere. 
    Results are shown for the $N=6$ spin-$\frac{1}{2}$ anisotropic $XY$ cyclic
    chain in a transverse magnetic field, $h=1$ 
    [see Eq.~(\ref{XY_H})] with $\gamma=0.5$, at points of interest (a) $\lambda=0$, 
    (b) $\lambda =1.15$, (c) $\lambda= 1.18$, and (d) 
    $\lambda=4$.  ${\rho}_{\mathrm{tot}}$ is shown again in larger 
    plots to enable better appreciation of the state of the system. 
    An animation showing all spin Wigner plots across the QPT is contained in 
    the supplementary material.
    The ground states for this Hamiltonian start similarly to those in Fig.~\ref{TISpinPlots}, where here (a) is also an eigenstate of $\hat{\sigma}_z$; resulting in a pure, separable state. This can be seen by considering the reduced states, especially the single-spin state.
    (b)-(d) show a steady loss in the purity (negative values) of the single-spin state as $\lambda$ increases.
    This, in addition to the full state being a pure state, is an indicator that the overall state is entangled.
    In the 2-, 3-, 4-, and 5-spin functions, we can see the spin-coherent state from $\lambda=0$ state split in two and move closer to the equator at orthogonal points on the Bloch sphere.
    At $\lambda=4$ the two coherent states are distinct and almost orthogonal -- it is clear that this approaches a GHZ-type state, which is the case as $\lambda \rightarrow \infty$.
    Note that (a) and (b) follow similarly to the transverse-field Ising model in Fig.~\ref{TISpinPlots}. After this point there is a parity change; similar to the difference between the GHZ states in Figs.~\ref{ref_plots}(g) and \ref{ref_plots}(h), the shift from the state in (b) to (c) shows a rotation of the interference terms around the $x$ axis.
    This results in a larger negative value at $\rho_{\text{tot}}(0,0)$, which is reflected in the phase transition in Fig.~\ref{XYPhaseLine}.
    Note that the correlations in $\rho_{12345}$ indicate that the state is not yet a perfect GHZ state, 
    for which all reduced states are equal statistical mixtures of $\ket{\uparrow}^{\otimes n}$ and $\ket{\downarrow}^{\otimes n}$.
    }
\end{figure*}

Figure~\ref{XYPhaseLine} shows the behavior of the $XY$ model as the coupling 
strength $\lambda$ is varied for a fixed value of the anisotropy parameter
$\gamma=0.5$, in terms of equal-angle slices 
of the GWFs. 
Phase lines for both the full system correlation function and the complete set 
of sub-system correlation functions (shown in Fig.~\ref{ChainCorrelationKey}) 
are shown. 

Abrupt changes in the ground state of the system are clearly seen from 
Fig.~\ref{XYPhaseLine} at both $\lambda=\lambda_{f}(\gamma=0.5) \approx 1.155$, which corresponds 
to the factorization point, and $\lambda\approx1.545$, which represents the 
pseudo-critical point in the $N=6$ system corresponding to the critical point
of the 2QPT at  $\lambda=\lambda_{c}=1$
in the infinite chain. Discontinuities at the 
factorization point $\lambda_{f}$ are seen in all correlation functions of 
the system, with higher-order correlations better capturing the transition. 
This highlights the fact that the factorization 
point coincides with an energy level crossing of the ground state and first 
excited state \cite{XY_QPT_1,XY_QPT_5,XY_QPT_5E}. 
The mean of the two values of $W^{\mathrm{EA}}(\theta,\phi)$ just before and just
after the factorization point is precisely that of an equal mixture of the two (positive
and negative) parity states $\ket{\vartheta_\pm}$.  Thus, the interference terms precisely
cancel there, giving the corresponding value of the state $\ket\vartheta$, viz.,
$W^{\mathrm{EA}}(0,0)=2^{-n}(1+\sqrt 3 \cos\vartheta)^n$, for any number $n\le N$ of spins. 
For the particular value $\gamma=\frac12$ that we have chosen to display in Fig.~\ref{XYPhaseLine},
Eq.~(\ref{fac-angle}) then yields the corresponding value $\vartheta=\cos^{-1}\frac1{\sqrt3}$,
leading to the fortuitous equality of the values of all of the correlation
functions at the factorization point in Fig.~\ref{XYPhaseLine}.

A second discontinuity is seen 
in many of the system correlation functions at $\lambda\approx1.545$, which coincides with the 2QPT 
pseudo-critical point. Higher-order correlations such as 
 $\rho_{\mathrm{tot}}$ and $\rho_{12345}$ experience abrupt jumps in their 
phase lines. Nearest-neighbor 
two-site and three-site correlation functions, $\rho_{12}$ and $\rho_{123}$ 
respectively, also present clear discontinuities in their phase 
lines, highlighting their utility for witnessing 2QPTs in larger 
systems \cite{XY_QPT_6, XY_QPT_4, XY_QPT_7}. Despite this, the 2QPT is less 
clearly seen in the $\rho_{1234}$ and $\rho_{135}$ correlation functions. 
The fact that $\rho_{1234}$ is unable to witness the 2QPT clearly is 
somewhat surprising as higher-order correlations such as the five- and six-site 
correlation functions present the clearest indications of the 2QPT.

Figure~\ref{XYSpinPlots} shows the spin Wigner plots at points of interest 
across the phase plot. Figure~\ref{XYSpinPlots}(a) shows the initial state 
of the system at $\lambda=0$. Here the system is in a state
with all spins aligned in the $z$ direction of the external field.
This is clearly reflected in all the spin 
Wigner plots as they match the aligned spin reference plots shown in 
Figs.~\ref{ref_plots}(a) and~(b). 

As $\lambda$ is increased, the total state of the system $\rho_{\mathrm{tot}}$ gradually transitions towards a state of the $\ket{\mathrm{GHZ}_x^+}$ type.
The lower-order correlations, while only able to capture some features of the GHZ state, display how the spins migrate from a coherent state oriented at the north pole and split into two coherent states aligned in the $z$--$x$ plane, moving in opposite directions.
For the $\lambda = 4$ plot in Fig.~\ref{XYSpinPlots}(d), we can see the two coherent states more clearly as they approach the $x$ axis, anti-aligned on opposite sides of the equator.
The higher-order correlations, such as $\rho_{12345}$ and $\rho_{\mathrm{tot}}$, clearly start to show oscillating positive and negative values -- characteristic features of a GHZ state. 
Note that the existence of correlations between the two coherent states in $\rho_{12345}$ shows that the state is not a perfect GHZ state, which only happens in the limit  $\lambda \rightarrow \infty$.

As with the transverse-field Ising model, it is clear that this is a gradual change to the $\ket{\mathrm{GHZ}_x^+}$ state; the difference here is at the factorization point $\lambda_{f}$ that causes an abrupt change in the symmetry of the system, which can be seen in Fig.~\ref{XYSpinPlots}(c). 
The state of the system then transitions abruptly from a positive-parity state to a negative-parity state.
These positive- and negative-parity states are simply the superposition of the two coherent states that are evident in the lower-order correlations, with either a positive phase or negative phase, which at the limit where the two states aligned along the $x$ axis is similar to the difference between a $\ket{\mathrm{GHZ}_x^+}$-type state and a $\ket{\mathrm{GHZ}_x^-}$-type state.  
This behavior is expected and has been documented in previous research \cite{XY_QPT_1,XY_QPT_5,XY_QPT_5E}. 

Finally, at $\lambda=\lambda_{c}$ we witness the system transition back to a positive-parity state as a critical point of the 2QPT is passed. 
The actual critical point of the system cannot be determined with any 
certainty as the factorization point obscures this area of the phase 
line. Further exploration of this phenomenon and its scaling 
properties as a function of system size could be fruitful.

\subsection{\label{XXZ}Spin-1/2 {\it XXZ} Spin Chain Model}

The Hamiltonian for a cyclic, spin-$\frac{1}{2}$ anisotropic 1D $XXZ$ Heisenberg model with 
nearest-neighbor interactions is given by
\begin{equation}
   \hat{\mathcal{H}} = \hat{\mathcal{H}}(J,\Delta) =
   \frac{J}{4} \sum_{i=1}^{N} \left[\hat{\sigma}^x_i 
   \hat{\sigma}^x_{i+1} + \hat{\sigma}^y_i \hat{\sigma}^y_{i+1} + \Delta 
   \hat{\sigma}^z_i \hat{\sigma}^z_{i+1}\right] \,,
   \label{XXZ_H}
\end{equation}
where $J$ is the coupling strength and $\Delta$ is the anisotropy parameter 
\cite{SpinChain1B,XY_QPT_7,XXZ_QPT_1}. As before, $\hat{\sigma}_i^\alpha,\,\alpha = x,y,z$ 
are the usual Pauli matrices, and periodic boundary conditions are assumed, 
so that $\hat{\sigma}^{\alpha}_{N+1} = \hat{\sigma}^{\alpha}_{1},\, \alpha =x,y,z$. 
Although the $XXZ$ Hamiltonian of Eq.~(\ref{XXZ_H})
contains two parameters, the essential physics is captured by just one, as we now show.

For simplicity we restrict attention to the case when $N$ is even and, 
by analogy with Eq.~(\ref{spin-parity-op}), 
we consider the unitary operator $\hat{U}_z$, defined as follows,
\begin{equation}
    \hat{U}_z \equiv (-i)^{\frac{N}{2}} \exp (i\pi\sum_{m=1}^{N /2}\frac{\hat{\sigma}_{2m}^z}{2}) = \prod_{m=1}^{N/2} \hat{\sigma}_{2m}^z\,.
\label{U_z-operator}    
\end{equation}
Considered as a similarity transformation, its mode of action is to perform
a rotation by $\pi$ about the global spin-space $z$ axis of the spins on the even sites,
under which $\hat{\sigma}_{2m}^\mu \rightarrow -\hat{\sigma}_{2m}^\mu\,, \mu = x,y\,; m=1,\dots,\frac{N}{2}$,
with all other spin components unchanged. Clearly, it leaves the underlying $SU(2)$ algebra unchanged.
However, its mode of action on the $XXZ$ Hamiltonian of Eq.~(\ref{XXZ_H}) is thus given by
\begin{equation}
    \hat{U}_z^{-1}\hat{\mathcal{H}}(J,\Delta)\hat{U}_z = \hat{\mathcal{H}}(-J,-\Delta)\,.
    \label{H_XXZ-sim-Xm}
\end{equation}
From Eq.~(\ref{H_XXZ-sim-Xm}) we immediately conclude that it is only the relative
sign between the parameters $J$ and $\Delta$ that is relevant, rather than their
separate signs.  Thus, with no real loss of generality we may take $J>0$ so long as we 
also consider $\Delta$ over the full range $-\infty < \Delta < \infty$.  Thus, henceforth
we put $J=+1$ to set the overall energy scale.

Only for the isotropic case when $\Delta=1$ does the $XXZ$ Hamiltonian of Eq.~(\ref{XXZ_H})
preserve the full $SU(2)$ symmetry of arbitrary rotations in spin space. However, it is easy to show 
that the Hamiltonian commutes with the $z$ component of the total spin operator,
\begin{equation}
    \hat{S}_T^z \equiv \frac{1}{2}\sum_{l=1}^N \hat{\sigma}_l^z\,,
    \label{S_T-z}
\end{equation}
which then leads to the smaller $U(1)$ symmetry under arbitrary rotations about the global
spin-space $z$ axis, which are generated by the rotation operator,
\begin{equation}
    \hat{R}_z(\phi) \equiv \mathrm{e}^{i\phi \hat{S}_T^z} = \prod_{l=1}^{N}\bigg[\cos(\frac{\phi}{2})\hat{\mathbbm 1}
    +i\sin(\frac{\phi}{2})\hat{\sigma}_l^z \bigg]\,.
    \label{rotation_z-op}
\end{equation}
The mode of action of $\hat{R}_z(\phi)$ on the basic spin operators is as follows,
\begin{equation}
    \begin{split}
        \hat{\sigma}_i^x & \to \hat{R}_z^{-1}(\phi) \hat{\sigma}_i^x \hat{R}_z(\phi) = 
        \cos{(\phi)}\,\hat{\sigma}_i^x + \sin{(\phi)}\,\hat{\sigma}_i^y \,,\\
        \hat{\sigma}_i^y & \to \hat{R}_z^{-1}(\phi) \hat{\sigma}_i^y \hat{R}_z(\phi) = 
        \cos{(\phi)}\,\hat{\sigma}_i^y - \sin{(\phi)}\,\hat{\sigma}_i^x \,,\\
        \hat{\sigma}_i^z & \to \hat{R}_z^{-1}(\phi) \hat{\sigma}_i^z \hat{R}_z(\phi) = \hat{\sigma}_i^z\,,
        \end{split} 
    \label{rotated-spin-ops}
\end{equation}
and it thus leaves the $XXZ$ Hamiltonian of Eq.~(\ref{XXZ_H}) invariant,
\begin{equation}
    \hat{R}_z^{-1}(\phi) \hat{\mathcal{H}}(J,\Delta) \hat{R}_z(\phi) = \hat{\mathcal{H}}(J,\Delta)\,.
    \label{H_XXZ-U(!)-invariant}
\end{equation}

To set the scene for the quantum spin-$\frac{1}{2}$ case, let us first consider the classical version 
of the $XXZ$ model Hamiltonian of Eq.~(\ref{XXZ_H}) with $J>0$, as used here, 
and for the limiting case $N\to\infty$.  For $\Delta<-1$ the
system is a ferromagnet aligned along either of the $\pm z$ spin-space directions.  At a first critical 
point, $\Delta_{c_{1}}=-1$, the system undergoes a first-order transition to a N{\'e}el
antiferromagnet with the axis of alignment in an arbitrary direction in the spin-space $x$--$y$ plane. 
This so-called easy-plane antiferromagnetic phase survives over the region $-1<\Delta<1$ until, at
the second critical point, $\Delta_{c_{2}}=1$, the system undergoes a further first-order transition
to a so-called easy-axis N{\'e}el antiferromagnet with the axis of alignment along the spin-space $z$
direction for $\Delta>1$.  The classical ground-state is thus doubly degenerate for $|\Delta|>1$ and
infinitely degenerate for $|\Delta|\leq 1$.

The spin-$\frac{1}{2}$ version of the $XXZ$ chain model Hamiltonian has been shown to be integrable
via the well-known Bethe Ansatz, for both finite and infinite values of $N$
\cite{SpinChain1B,Bethe_1931,Orbach_1958,Walker_1959,Yang-Yang_1,Yang-Yang_2,Johnson-McCoy_1972,Baxter_1973}. 
We now discuss the exact solution for the system in the thermodynamic limit ($N\to\infty$).
It transpires that although the quantum spin-$\frac{1}{2}$ version of the model in this limit maintains the
 same two critical points as the classical version, the nature of two of the phases 
 and the transition at $\Delta_{c_{2}}=1$ are substantially different 
 \cite{SpinChain1B,Bethe_1931,Orbach_1958,Walker_1959,Yang-Yang_1,Yang-Yang_2,Johnson-McCoy_1972,Baxter_1973,XY_QPT_7,XXZ_QPT_1,XXZ_QPT_2,XXZ_QPT_3}.  
 
 By writing the first two terms in the Hamiltonian of Eq.~(\ref{XXZ_H}) in the form
 $\hat{\sigma}^x_i \hat{\sigma}^x_{i+1} + \hat{\sigma}^y_i \hat{\sigma}^y_{i+1}=
 \hat{\sigma}^+_i \hat{\sigma}^-_{i+1} + \hat{\sigma}^-_i \hat{\sigma}^+_{i+1}$, where
 $\hat{\sigma}^\pm_j \equiv \frac{1}{\sqrt{2}}(\hat{\sigma}^x_j\pm i\hat{\sigma}^y_j)$ are the usual
 Pauli spin raising and lowering operators, it is easy to see that the two spin-space
 $z$-aligned ferromagnetic states are always exact eigenstates of this Hamiltonian. 
 By contrast, we also see at once that the corresponding $z$-aligned N{\'e}el
antiferromagnetic states are {\it not\,} eigenstates.
 For all values $\Delta<-1$ these two ferromagnetic states form the doubly-degenerate ground state, just 
 as in the classical case.  The order parameter is the absolute value of the $z$ component of the 
 magnetization, which assumes its maximally saturated value $|M^z|=\frac{1}{2}$ throughout the region
 $\Delta<-1$, over which the energy spectrum is gapped.
 
 Precisely at the value $\Delta=-1$, the order parameter $M^z$ drops to zero discontinuously,
 and we have a 1QPT. Throughout the ensuing region $-1<\Delta<1$
 the spectral threshold is also gapless.  The quantum fluctuations completely destroy the classical
 easy-plane N{\'e}el order that exists there.  Indeed, there is {\it no} long-range 
 order present in the spin-$\frac{1}{2}$ model
 in this region, and accordingly the ground state also becomes non-degenerate there.  Such ground-state phases 
 that are completely devoid of all long-range order are called {\it critical phases}. 
 We note that this critical phase also contains the very special point $\Delta=0$, which
 can be exactly mapped by a Jordan-Wigner transformation onto free (lattice) fermions.
 
 In this planar-regime critical phase with $-1<\Delta<1$
 all equal-time two-spin correlation functions decay algebraically with separation distance,
 with an exponent that depends on the value of $\Delta$.
 It is only at the exact limiting value $\Delta=-1+\epsilon$ (with $\epsilon$ a positive infinitesimal)
 that the ground state again breaks the symmetry of the Hamiltonian. It is a doubly-degenerate state
 with a fully saturated value of the {\it staggered\,} magnetization 
 (i.e., as defined on either of the two sublattices of even or odd sites) pointing in
 a direction perpendicular to the spin-space $z$ axis, and with no correlated fluctuations.
 {\it A priori\,}, we do not expect to see features of such critical phases in small finite-sized
 systems far removed from the thermodynamic limit ($N\to\infty$) \cite{QuantPhaseTrans, QuantPhaseTrans2}.
 
 At $\Delta_{c_{2}}=1$, the first-order classical transition now becomes a $\infty$QPT
 of the Berezinskii–Kosterlitz–Thouless type, and 
 the ground state reverts to being doubly degenerate, such that for $\Delta>1$ antiferromagnetic
 long-range order gradually develops as $\Delta$ increases.  The (N\'eel) order parameter for the region
 $\Delta>1$ is the staggered magnetization, $m^z$, in the spin-space z direction, which was first
 calculated by Baxter \cite{Baxter_1973}. He showed explicitly that $m^z$ develops smoothly 
 and monotonically as a function of $\Delta$ from a value equal to zero for $\Delta=1$, only reaching
 its fully saturated (classical) value in the limit $\Delta\to\infty$. Similarly, the whole region
$\Delta>1$ has an energy threshold that is gapped, with an energy gap that also increases smoothly
and monotonically from zero at $\Delta=1$ as $\Delta$ is increased further.
 
As previously stated, the finite size ($N=6$) of our system enables exact solutions 
of the ground state to be calculated through direct diagonalization. 
In turn, the system's density matrix and complete set of reduced density 
matrices (see Fig.~\ref{ChainCorrelationKey}) can then be calculated. 

\begin{figure}
    \includegraphics[width=\linewidth]{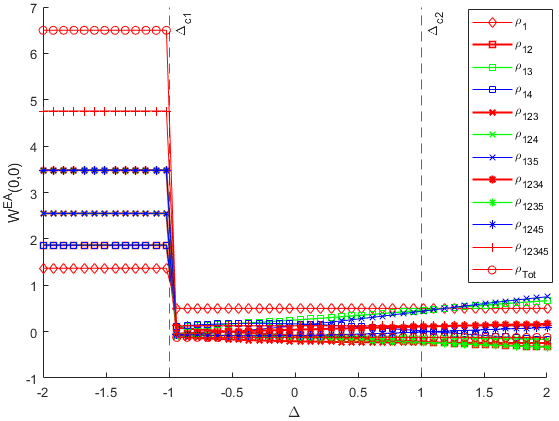}
    \caption{\label{XXZPhaseLine} Equal-angle slices of the reduced GWFs,
    $\rho_{I}(\theta,\phi) \equiv W^{\mathrm{EA}}_{I}(\theta,\phi)$, 
    shown in Fig.~\ref{ChainCorrelationKey}, taken at $\theta=\phi=0$, 
    for  the $N=6$ spin-$\frac{1}{2}$ $XXZ$ anisotropic cyclic Heisenberg chain
    [see Eq.~(\ref{XXZ_H})] with $J=1$. The 1QPT critical point 
    $\Delta_{c_{1}}$ and the $\infty$QPT critical point $\Delta_{c_{2}}$ are marked 
    on the plot. A full sized version of this plot is available in the 
    supplementary material.
    }
\end{figure}
    
\begin{figure*}
    \includegraphics[width=\textwidth]{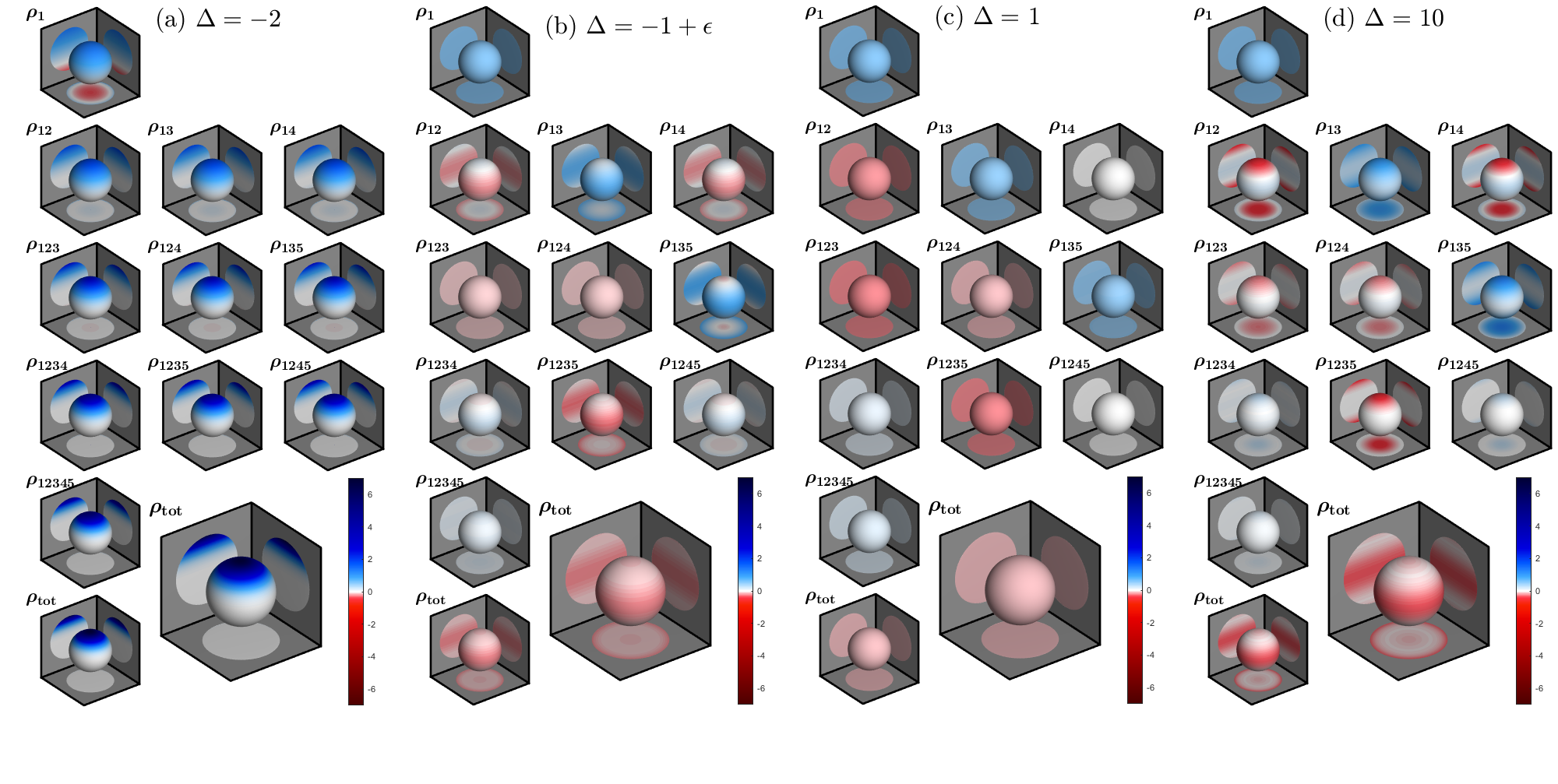}
    \caption{\label{XXZSpinPlots} Equal-angle slices of the reduced GWFs,
    $\rho_{I}(\theta,\phi) \equiv W^{\mathrm{EA}}_{I}(\theta,\phi)$, 
    shown in Fig.~\ref{ChainCorrelationKey}, plotted on the Bloch sphere.
    Results are shown for the $N=6$ 
    spin-$\frac{1}{2}$ $XXZ$ anisotropic cyclic Heisenberg chain 
    [see Eq.~(\ref{XXZ_H})] with $J=1$, at points of interest (a) $\Delta=-2$, 
    (b) $\Delta=\Delta_{c_{1}}+\epsilon=-1+\epsilon$ 
    (where $\epsilon$ is a positive infinitesimal), (c) $\Delta=\Delta_{c_{2}}=1$, and 
    (d) $\Delta=10$. 
    ${\rho}_{\mathrm{tot}}$ is shown again in larger plots 
    to enable better appreciation of the state of the system. An animation 
    showing all spin Wigner plots across the QPT is contained in the 
    supplementary material.
    As with the other spin chains, the first ground state is the pure, separable state where 
    all spins are aligned along the $z$ axis, which can be seen in (a).
    This model differs in that the ground state, ignoring the degeneracy, stays as this state until the phase transition at $\Delta = -1$, this is also evident from Fig.~\ref{XXZPhaseLine}, where all the correlation functions are constant until the phase transition. 
    (b) shows the GWFs for the state just after the phase transition; it demonstrates how the ground state after the phase transition becomes highly entangled.
    This is clear from noting that the single-qubit state is a fully mixed state while the full six-spin function is a pure state. 
    Note that all subsequent states also have a fully mixed state in the single-spin Wigner function. 
    They then all differ in how the entanglement is distributed throughout the state, which can be seen from the 2-, 3-, and 4-spin reduced Wigner functions.
    For instance, in (c) we have a completely uniform distribution in all equal-angle slices. 
    For $\Delta=1$ all equal-angle GWF are isotropic as the Hamiltonian is invariant under spin rotation and has a singlet ground state. The two-spin cases $\rho_{1k}$ are Werner states.
    In (b) and (d) there are more features evident in the reduced states. This is because they also have contributions from the triplet state $\ket{\Psi_+}$ shown in Fig.~\ref{ref_plots}(d).
    As $\Delta$ increases, in (d) we can see in some of the reduced states, such as $\rho_{13}$ and $\rho_{135}$, that this is similar to a GHZ-state state in that there are two antipodal coherent states on the Bloch sphere. 
    We can further see that that the full state tends towards that in Fig.~\ref{ref_plots}(j).
    }
\end{figure*}

Figure~\ref{XXZPhaseLine} shows the behavior of the $XXZ$ model as 
the anisotropy parameter $\Delta$ is varied. Once again the equal-angle 
slice of the GWF has been calculated for the full system correlation function 
and the constituent subsystem correlation functions. Spin Wigner functions 
are plotted at points of interest across the phase diagram in 
Fig.~\ref{XXZSpinPlots}.

A clear discontinuity is seen in the phase lines at the 1QPT critical point 
$\Delta_{c_{1}}$, with all of the system's correlation functions exhibiting 
an abrupt jump. We note parenthetically that if the symmetry had been 
preserved by using the state $\ket{\mathrm{GHZ}_z^+}$ 
with an equal mixture of ferromagnetic up and down states,
the discontinuity in $\rho_1$ at $\Delta_{c_1}$ would presumably disappear.
In general, higher-order correlations are better able to capture the 
behavior of the system at this critical point. Although past research was 
unable to witness 1QPT critical behavior in the single site GWF 
\cite{Discrete}, in our findings a clear discontinuity at the 1QPT critical 
point is seen in ${\rho}_{1}$. An explicit maximum or minimum at the $\infty$QPT 
critical point, $\Delta_{c_{2}}$, is not seen however. 

Despite this, 
interesting behavior is captured in many of the system's correlation 
functions across the $-1<\Delta<10$ range. This type of behavior has been 
captured in past research \cite{Discrete}. ${\rho}_{1}$, ${\rho}_{12345}$ and 
${\rho}_{\mathrm{tot}}$ all take constant values across the range.
${\rho}_{12}$ experiences a decrease in the value of the GWF across the range. 
The first derivatives of ${\rho}_{13}$ and ${\rho}_{135}$ experience a minimum 
at $\Delta\approx1$, while ${\rho}_{124}$ and ${\rho}_{1235}$ both experience 
a maximum in their first derivative at this point. This highlights how 
exploration of the first derivatives of the GWF can be used to witness 
the critical point of a $\infty$QPT. ${\rho}_{14}$ and ${\rho}_{124}$ both experience a maximum at $\Delta\approx-0.3$, while ${\rho}_{1245}$ experiences a minimum 
at $\Delta\approx-0.3$.  The first derivatives of ${\rho}_{123}$ and 
${\rho}_{1234}$ decrease across the range. It has been shown that through 
extremization procedures certain correlation measures can better witness $\infty$QPTs 
\cite{XY_QPT_1, XY_QPT_7, XXZ_QPT_4, XXZ_QPT_5}. Past research has 
implemented these for the GWF with much success \cite{Discrete}, and such 
further exploration of these techniques could be useful.

At $\Delta_{c_{1}}$ we have a 1QPT between a non-degenerate ground state and a two-dimensional 
ground-state space. Unlike in the case of the transverse Ising model, there is no adiabatic continuity criterion 
to select a vector in this space,
so, for example, neither the state $\ket{\uparrow\uparrow\cdots}$ with 
all spins aligned in the upwards $z$ direction nor the state
$\ket{\mathrm{GHZ}_z^+} \equiv\frac{1}{\sqrt{2}}(\ket{\uparrow\uparrow\cdots}+\ket{\downarrow\downarrow\cdots})$ 
has any special claim.  The unbiased choice for $\Delta<-1$ is presumably then a mixture $\hat{\rho}
=\frac{1}{2}(\ket{\uparrow\uparrow\cdots}\bra{\uparrow\uparrow\cdots}+\ket{\downarrow\downarrow\cdots}\bra{\downarrow\downarrow\cdots})$.
This will have the same reduced density matrices as the state $\ket{\mathrm{GHZ}_z^+}$.
Nevertheless, for simplicity Fig.~\ref{XXZSpinPlots} shows the state $\ket{\uparrow\uparrow\cdots}$.
Thus, Fig.~\ref{XXZSpinPlots}(a) clearly shows the state 
of the system at $\Delta=-2$ to be in an aligned ferromagnetic state; this can 
be inferred due to the resemblance of the state's spin Wigner plots to 
the reference plots of Fig.~\ref{ref_plots}(l) for the ferromagnetic mixed state.

Figure~\ref{XXZSpinPlots}(b) shows the state 
of the system at the 1QPT critical point $\Delta=-1+\epsilon$. The state of 
${\rho}_{\mathrm{tot}}$ and ${\rho}_{1235}$ show similarities to the 
entangled, antiparallel 
aligned state seen in Fig.~\ref{ref_plots}(j), while the state of ${\rho}_{1234}$ 
and ${\rho}_{1245}$ present similarities to the anti-aligned state in 
Fig.~\ref{ref_plots}(c). On the other hand, ${\rho}_{123}$ and ${\rho}_{124}$ are both in a low
amplitude singlet-type state [see Fig.~\ref{ref_plots}(e)], while ${\rho}_{135}$ 
and ${\rho}_{13}$ both show similarities to the Bell state of 
Fig.~\ref{ref_plots}(d). The states of ${\rho}_{12}$ and ${\rho}_{14}$ possess similarities 
to the anti-aligned entangled state in Fig.~\ref{ref_plots}(f). The abrupt 
emergence of these states at the 1QPT critical point enable us to witness 
the 1QPT and infer the presence of an antiferromagnetic type phase.

Figure~\ref{XXZSpinPlots}(c) shows the state of the system at the $\infty$QPT 
critical point $\Delta=1$.  Here all equal-angle Wigner functions are constant over the Bloch sphere as the Hamiltonian is an isotropic antiferromagnet. The signs of the two-spin reduced Wigner functions show that the nearest-neighbor correlation (represented by $\rho_{12}$) is negative. 

The states that have a constant value over the Bloch sphere for the equal-angle slice in Fig.~\ref{ref_plots} are the mixed state (k) and the singlet state (e), that has a constant value of $-\frac{1}{2}$.
The two-spin maximally mixed state, $\frac{1}{4}\mathbbm{1}$, yields similar results to the single-spin case with a constant value of $\frac{1}{4}$ everywhere.
The two-spin states in Fig.~\ref{XXZSpinPlots}(c) are statistical mixtures of these two states and are the Werner states~\cite{Werner}
\begin{equation}
    \hat{\rho}(x) = x\ket{\Psi_-}\bra{\Psi_-}+\frac{(1-x)}4 \mathbbm 1,
\end{equation}
where $\ket{\Psi_-} = \frac{1}{\sqrt{2}}(\ket{\uparrow\downarrow} - \ket{\downarrow\uparrow})$ and $x\in \left[0,1\right]$; note that $\hat{\rho}(x)$ is only entangled when $x > \frac13$~\cite{Werner}.

The equal-angle slice for the Wigner function of a Werner state is 
\begin{equation}
    W(\theta,\phi; x)=\frac{1-3x}4,
\end{equation}
and is only an entangled state when the equal-angle slice has a negative value.
We see in Fig.~\ref{XXZSpinPlots}(c) that the value of the Wigner function changes as the distance between the spins changes, with explicit values of
\begin{equation}
\begin{split}
    \rho_{12} &= \frac{1-\sqrt{13}}{12} \approx -0.217, \; x = \frac{2+\sqrt{13}}{9} \approx 0.623,\\
    \rho_{13} &= \frac14 + \frac{3\sqrt{13}}{52} \approx 0.458, \; x = -\frac{1}{\sqrt{13}} \approx -0.277, \\
    \rho_{14} &= \frac{2\sqrt{13}}{39} -\frac16 \approx 0.018, \; x =\frac59 -\frac{8\sqrt{13}}{117}\approx 0.309.
\end{split}
\end{equation} 

Note that on both sides of the $\Delta = 1$ critical point, the two-spin states include the statistical mixture of the $\ket{\Psi_+} = \frac{1}{\sqrt{2}}(\ket{\uparrow\downarrow} + \ket{\downarrow\uparrow})$ from Fig.~\ref{ref_plots}(d), resulting in a general state
\begin{equation}\label{doubleWernerState}
    \hat{\rho}(x,y) = x\ket{\Psi_+}\bra{\Psi_+} + y\ket{\Psi_-}\bra{\Psi_-}+\frac{(1-x-y)}{4}\mathbbm{1}.
\end{equation}
This is what gives rise to the $\theta$ dependence in the equal-angle slice, however it is worth noting 
that by adding this term, negativity in the Wigner function no longer holds as a measure of entanglement.

Similarly, for the three-spin functions, we have a statistical mixture of the mixed state and 
singlet states between the three spin pairs, resulting in an overall state
\begin{equation}
    \hat{\rho}(x,y,z) = x\Psi^-_{12} + y\Psi^-_{23} + z\Psi^-_{13} + \frac{(1-x-y-z)}{8}\mathbbm{1},
\end{equation}
where $\Psi^-_{12} =  \ket{\Psi_-}\bra{\Psi_-} \otimes \frac{1}{2}\mathbbm{1}$, and the other subscripts 
indicate the coupling between the other pairs of spins.
Note that the Wigner function for  $\Psi^-_{12}$ is 
$\Trace[(\ket{\Psi_-}\bra{\Psi_-} \otimes \frac{1}{2}\mathbbm{1}) (\Delta_1 \otimes \Delta_2)] = \frac{1}{2}\Trace\left[\ket{\Psi_-}\bra{\Psi_-}\Delta_1 \right] \Trace\left[\Delta_2 \right] = -\frac{1}{4}$.
Further, the Wigner function for all $\Psi^-_{ij}$ yields the same result.
The Wigner function for $\hat{\rho}(x,y,z)$ therefore has a constant value 
\begin{equation}
    W(\theta,\phi; x,y,z) = \frac{1-3(x+y+z)}{8}\,.
\end{equation}
In our case the three three-spin correlation functions shown in Fig.~\ref{ChainCorrelationKey}
can be explicitly evaluated as follows,
\begin{equation}
\begin{split}
\rho_{123} &= -\frac{1}{24} - \frac{17\sqrt{13}}{312} \approx -0.238, \text{ where, } \\
& x=\frac29 + \frac{\sqrt{13}}{9},\; y =\frac29 + \frac{\sqrt{13}}{9},\; z=-\frac{1}{\sqrt{13}} \\
\rho_{124} &= -\frac16 + \frac{\sqrt{13}}{78} \approx  -0.120, \text{ where, }\\
& x=\frac29 + \frac{\sqrt{13}}{9},\; y=- \frac{1}{\sqrt{13}},\; z=\frac59 - \frac{8\sqrt{13}}{117}; \\
\rho_{135} &= \frac18 + \frac{9\sqrt{13}}{104} \approx 0.437, \text{ where, }\\
& x=- \frac{1}{\sqrt{13}} ,\; y=- \frac{1}{\sqrt{13}}, \; z=- \frac{1}{\sqrt{13}},
\end{split}
\end{equation}

The two-spin states in Figs.~\ref{XXZSpinPlots}(b) and~(d) have a similar structure to those in Figure~\ref{XXZSpinPlots}(c), although we now see a $\theta$ dependence in the Wigner function. 
This comes from the additional $\ket{\Psi_+})$ term discussed above, 
the pattern of which can be inferred from Fig.~\ref{ref_plots}(d).
Like the general two-spin reduced state from the ground states of this Hamiltonian (for $\Delta>-1$) are of the form in 
Eq.~(\ref{doubleWernerState}), the three-spin reduced states also have a 
contribution from $\Psi^+_{ij}$ (defined similarly to $\Psi^-_{ij}$ above),
which contribution goes to zero when $\Delta = 1$.


Finally, Fig.~\ref{XXZSpinPlots}(d) shows the state of the system at $\Delta=10$.  
${\rho}_{\mathrm{tot}}$ becomes less uniform in its value and tends towards the state shown in Fig.~\ref{ref_plots}(j), $\frac{1}{\sqrt{2}}\left( \ket{\uparrow\downarrow\uparrow\downarrow\uparrow\downarrow} - \ket{\downarrow\uparrow\downarrow\uparrow\downarrow\uparrow} \right)$.
Evidence for this type of state can further be seen from the reduced functions.

We saw in the other two Hamiltonians that as the state turns to a GHZ-type state, 
the reduced states show evidence of two antipodal coherent states, pointing in 
orthogonal directions on the Bloch sphere.
A similar effect is seen here. 
For $\rho_{13}$ and $\rho_{135}$, we have odd-number indices for the spins, 
picking out every other spin from the full state. 
This, in effect, gives us an equal statistical mixture of $\ket{\uparrow}$ 
and $\ket{\downarrow}$ states.
We can then see in the Wigner functions two coherent states, one on the north pole 
and the other on the south pole; the two- and three-spin equivalents to Fig.~\ref{ref_plots}(l).

In $\rho_{12}$ and $\rho_{14}$ we instead pick out anti-aligned features from the full state.
We therefore tend towards an equal statistical mixture of $\ket{\uparrow\downarrow}$ 
and $\ket{\downarrow\uparrow}$ in both cases.
In the equal-angle slice, both $\ket{\uparrow\downarrow}$ and $\ket{\downarrow\uparrow}$ 
have the same Wigner function representation.
In addition, the Wigner function of the statistical mixture of two states is the 
same as the addition of the two Wigner functions.
This statistical mixture then produces a Wigner function reminiscent of 
Fig.~\ref{ref_plots}(c) for $\rho_{12}$ and $\rho_{14}$.
Similar results are seen in the three-spin functions, $\rho_{123}$ and $\rho_{124}$ 
where we have a statistical mixture of a state with two $\ket{\uparrow}$ and 
one $\ket{\downarrow}$ with a state that has one $\ket{\uparrow}$ and two $\ket{\downarrow}$ spins.

As we move onto the four-spin states, the same effect can be seen, 
where we have the equal statistical mixture of $\ket{\uparrow\downarrow\uparrow\downarrow}$ and $\ket{\downarrow\uparrow\downarrow\uparrow}$ in $\rho_{1234}$ and $\rho_{1245}$.
Conversely, $\rho_{1235}$ differs as it doesn't have the same number of $\ket{\uparrow}$ and $\ket{\downarrow}$ spins in each side, which can been seen in the spin indices in the full state.

This all then paints a picture that we have a GHZ-type state that has every other spin flipped around the Bloch sphere, further indicating that the system is in an antiferromagnetic phase at $\Delta=10$. 
This behavior has been recorded in past research \cite{XXZ_QPT_1, XXZ_QPT_2}. 
In contrast to phase line techniques, spin Wigner function visualizations 
were able to witness the $\infty$QPT exactly at the critical point without the use 
of extremization procedures. This highlights the method's utility in 
witnessing features of a system that are less easily discerned through 
other techniques.

\section{\label{conclusion}Conclusions}

We have demonstrated the utility of the generalized Wigner function formalism in 
exploring quantum critical behavior in a range of spin chain models. 
The generalized Wigner function's ability to explore the ground state of many-body 
systems, while also capturing the quantum correlations present within them, 
has been shown to enable first-, second- and infinite- order quantum phase transitions 
to be witnessed and characterized in a range of systems. In addition to this, 
its generalization to discrete, finite-sized spin systems has been shown to allow 
for exhaustive exploration of each model's correlation functions, which in turn 
has enabled a much deeper appreciation of the critical behavior of these spin systems. 
We also demonstrated the GWFs ability to determine the state of the system 
through the use of spin Wigner function visualization techniques and qubit 
state reference plots. These reference plots enable us reliably to infer 
the state of the system through observation alone, making this method much 
more intuitive and accessible than other leading techniques. In addition to this, 
the spin Wigner plots highlight important features of second- and infinite-order 
QPTs that fixed angle phase lines are either unable or less successful at capturing. 

To further explore the utility of this method for witnessing second- and 
infinite-order QPTs, extremization of the GWF through careful choice of angles 
($\theta,\phi$) should be explored, as this has shown to be useful in 
characterizing higher order QPTs  \cite{GWF2,XY_QPT_1, XY_QPT_7, XXZ_QPT_4, XXZ_QPT_5}. 
The properties of ground-state factorization were also explored in the 
$XY$ model with the phenomena being captured in a wider range of correlation functions 
than past research in this area~\cite{Discrete}. We were able to capture 
both the entanglement transition and the characteristic change in symmetry 
of the system through the GWF alone. 

One interesting area of inquiry left 
to future research is the obscuring of the 2QPT critical point by the 
$XY$ model's ground-state factorization. The length over which the state 
of the system remains in the factorized state and its scaling with system size 
should be studied. This work also highlights the continued utility of finite 
spin systems in exploring quantum critical behavior as, despite their simplicity, 
they are able to produce a rich and insightful look into critical spin chains 
and many-body systems as a whole  
\cite{XY_QPT_1,XY_QPT_2,XY_QPT_3,XY_QPT_7,TI_QPT_1}. In addition to this, 
scaling of finite critical quantum systems is well captured by our method, thereby 
allowing for the properties of this phenomenon to be explored
further. The ability of phase-space techniques to provide in-depth and insightful 
analyses of many-body systems highlights their utility for better understanding 
more exotic states of matter present in this field of study  \cite{Exotic1,Exotic2,Exotic3}. 
Our method also provides a foundation upon which future research can apply 
phase-space techniques to discern the factors that lead to the emergence of different 
states and phenomena in these systems. The fact that the Wigner function 
can be reconstructed from experimental data highlights this method for 
implementation in the experimental investigation of quantum phase 
transitions \cite{Wigner2}.  

On a final note, is is interesting to compare the current GWF formalism with other widely
used quantum many-body theory techniques that also have a similar phase-space underpinning.
It is widely acknowledged that the coupled cluster method (CCM) \cite{Bishop_1991,Bishop_1998} nowadays 
provides one of the most pervasive, most powerful, and most accurate at attainable levels of 
implementation of all fully microscopic formulations of quantum many-body theory.  Thus, the CCM (in 
both its so-called normal and extended versions) provides an extremely convenient parametrization
of the Hilbert space, which also {\it exactly\,} maps it onto a classical Hamiltonian quantum many-body/field
theory in a complex symplectic phase space \cite{Arponen-Bishop-Pajanne_1987,Arponen-Bishop_1991,Arponen-Bishop_1993a,Arponen-Bishop_1993b}. 

In CCM formalisms one constructs explicitly an energy functional that variationally determines both the
ground-state wave function and the dynamical equations of motion for non-stationary states.  The equations
of motion thus take the familiar classical canonical Hamiltonian form in some well-defined many-body
configuration space.  The expectation-value functional of an {\it arbitrary\,} operator is also
manifestly constructed in a way that, very importantly, automatically preserves the Hellmann-Feynman
theory exactly at all natural levels of truncation in the configuration space \cite{Bishop-Arponen_1990,Bishop_1998}.  
Another key feature of the CCM techniques is that they also automatically satisfy the Goldstone linked-cluster
theorem at all levels of implementation \cite{Bishop_1991,Bishop_1998}, so that the thermodynamic 
limit ($N\to\infty$) can be taken from the very outset, thereby obviating the need for any finite-size
scaling of the results, as is needed in most other many-body calculations using alternative techniques.

In the present context we note that the CCM has been very extensively applied with great success to a plethora of systems
in quantum magnetism, at high levels of implementation in a systematic hierarchy of approximations
that are tailor-made for systems confined to a regular spatial lattice in any number of dimensions 
\cite{Farnell-Bishop_2004}.  Among many others, it has been applied in particular to the transverse Ising
model \cite{Farnell-Bishop-Richter_2019}, the $XY$ model \cite{Farnell-Kruger-Parkinson_1997}, and the 
$XXZ$ model \cite{Bishop-Farnell-Parkinson_1996,Bishop-et-al_2000}, both for the spin-$\frac{1}{2}$ chains
considered here as well as for higher spin values and on specific lattices in two or more dimensions. It will 
surely be of considerable future interest to explore possible synergies between the CCM and the current
GWF techniques in these and other applications to spin-lattice problems.

One particular area of such mutual interest is to extend the discussion to the {\it dynamical\,} phase
transitions \cite{Sciolla-Biroli_2011} that can occur when a quantum many-body system is driven out of
equilibrium, e.g., by a quantum quench.  After such a quench the long-time steady state can display a 
symmetry-broken phase, with singular features at the transition to the disordered phase.  If the energy
of the system is thereby shifted across a symmetry-restoration threshold and the system thermalizes, such
a transition may be thought of as occurring in the microcanonical ensemble. Alternatively, {\it non-ergodic\,}
systems can display long-time steady states that are unable to be described via any of the conventional
ensembles of statistical mechanics.  In such cases one can generate phases and phase transitions that are
unable to be found in any equilibrium situation \cite{Huse-et-al_2013}.

Spin-$\frac{1}{2}$ transverse-field Ising models provide a paradigmatic example of such 
systems \cite{Sciolla-Biroli_2011}, the dynamical phase transitions displayed by which have recently 
been studied \cite{Khasseh-et-al_2020} by means of a discrete truncated Wigner approximation \cite{Wigner4}, 
which shares some features with our own GWF formalism utilized here.  It would hence surely be of
considerable interest to investigate such dynamical phase transitions using the techniques discussed here,
possibly in conjunction with additional input inspired by the CCM.

\section*{Acknowledgments}
One of us (RFB) gratefully acknowledges the
Leverhulme Trust (United Kingdom) for the award of an
Emeritus Fellowship (EM-2020-013).
RPR acknowledges support from EPSRC grant numbers EP/N509516/1 and EP/T001062/1.


\bibliography{main_bib}

\end{document}